\documentclass[12pt]{article}
\usepackage{epsfig}
\newlength{\dinwidth}                                                           
\newlength{\dinmargin}                                                          
\catcode`@=11

\def\@citex[#1]#2{\if@filesw\immediate\write\@auxout{\string\citation{#2}}\fi
  \def\@citea{}\@cite{\@for\@citeb:=#2\do
    {\@citea\def\@citea{,\penalty\@m}\@ifundefined
      {b@\@citeb}{{\bf ?}\@warning
       {Citation `\@citeb' on page \thepage \space undefined}}%
\hbox{\csname b@\@citeb\endcsname}}}{#1}}

\def\citer{\@ifnextchar [{\@tempswatrue\@citexr}{\@tempswafalse\@citexr[]}}

\def\@citexr[#1]#2{\if@filesw\immediate\write\@auxout{\string\citation{#2}}\fi
  \def\@citea{}\@cite{\@for\@citeb:=#2\do
    {\@citea\def\@citea{--\penalty\@m}\@ifundefined
       {b@\@citeb}{{\bf ?}\@warning
       {Citation `\@citeb' on page \thepage \space undefined}}%
\hbox{\csname b@\@citeb\endcsname}}}{#1}}
\catcode`@=12

\def\a{\alpha}
\def\b{\beta}

\def\d{\delta}
\def\e{\epsilon}
\def\f{\phi}
\def\g{\gamma}
\def\h{\eta}

\def\j{\psi}

\def\m{\mu}

\def\o{\omega}
\def\p{\pi}

\def\s{\sigma}

\def\x{\xi}

\def\G{\Gamma}

\def\L{\Lambda}

\def\co{{\cal O}}

\def\bo{{\raise.15ex\hbox{\large$\Box$}}}               
\def\face{{\raise.2ex\hbox{$\displaystyle \bigodot$}\mskip-2.2mu \llap {$\ddot
        \smile$}}}                                      

\def\Bar#1{\overline{#1}}                       
\def\leftrightarrowfill{$\mathsurround=0pt \mathord\leftarrow \mkern-6mu
        \cleaders\hbox{$\mkern-2mu \mathord- \mkern-2mu$}\hfill
        \mkern-6mu \mathord\rightarrow$}       
\def\dvec#1{\vbox{\ialign{##\crcr
        \leftrightarrowfill\crcr\noalign{\kern-1pt\nointerlineskip}
        $\hfil\displaystyle{#1}\hfil$\crcr}}}           

\def\beq{\begin{equation}}
\def\eeq{\end{equation}}

\def\beqx{\begin{displaymath}}
\def\eeqx{\end{displaymath}}

\def\beql{\begin{eqnarray}}
\def\eeql{\end{eqnarray}}
\def\NO{\nonumber}
\def\msb{\overline{\rm MS}}

\def\Journal#1#2#3#4{{#1}{\bf #2} (#4) #3}

\def\NPB{{\em Nucl. Phys.} {\bf{B}}}
\def\PLB{{\em Phys. Lett.} {\bf{B}}}
 
\def\PRD{{\em Phys. Rev.} {\bf{D}}}
\def\ZPC{{\em Z. Phys.} {\bf{C}}}

\begin{document}

\begin{flushright}
NIKHEF/98-025\\
September 1998
\end{flushright}

\vspace{15mm}
\begin{center}
{\Large\bf\sc Soft Gluon Resummation for\\ }
\vspace{5mm}
{\Large\bf\sc Heavy Quark Electroproduction}
\end{center}
\vspace{2cm}
\begin{center}
{\large Eric Laenen and Sven-Olaf Moch}\\
\vspace{15mm}
{\it NIKHEF Theory Group\\
P.O. Box 41882, 1009 DB Amsterdam, The Netherlands} \\
\end{center}

\vspace{2cm}

\begin{abstract}

We present the threshold resummation for the cross section
for electroproduction of heavy quarks. 
We work to next-to-leading logarithmic accuracy, and in single-particle 
inclusive kinematics. 
We provide next-to-leading and next-to-next-to-leading order 
expansions of our resummed formula, and
examine numerically the quality of these finite order approximations.
For the case of charm we
study their impact on the structure function $F_2$ and its 
differential distribution with respect to the charm transverse momentum.

\end{abstract} 

\thispagestyle{empty}

\newpage \setcounter{page}{2}

\section{Introduction}

The production of heavy quarks in deep-inelastic 
scattering is a reaction of great 
interest because it sheds light on a number of fundamental 
issues in the QCD-improved parton model. First, at moderate 
momentum transfer $Q$, it is a direct probe of the gluon density in the 
proton. Second, it allows the exploration of the relevant degrees of freedom, as a 
function of $Q$, for describing this process,
the options comprising the treatment of the heavy quark as 
a quantum field external to the proton, via an ``effective'' 
parton density, or as a combination thereof. Finally, detailed studies of 
perturbative QCD dynamics are possible using the presence
of at least two hard scales: $Q$ and the heavy quark mass $m$.

Charm quarks have been identified in deep-inelastic 
electron proton collisions by the EMC collaboration \cite{EMC}
at small $Q$ and low final state invariant mass (large $x$),
and more recently and in much greater numbers at high $Q$ 
and medium to small $x$ by the ZEUS and H1 collaborations at HERA \cite{HERAc}.
Considerably more charm (and bottom) data are anticipated at HERA.

At the theoretical level the reaction has already been studied extensively.
In the framework where the heavy quark is not treated as a 
parton, leading order (LO) \cite{gr} and next-to-leading order 
(NLO) \cite{lrsvn93} calculations of the inclusive structure functions
exist. Moreover, LO \cite{is} and NLO \cite{hs} 
calculations of fully differential distributions have 
been performed in recent years.
So far theory agrees reasonably well with the HERA data.
We shall not concern ourselves here with descriptions
in which the heavy quark is (partly) treated as a parton
\cite{vfns}.

At the HERA center of mass energy of 314 GeV,
one naively expects
the production of a heavy quark pair with a minimum invariant mass of 
a couple of GeV 
to be insensitive to threshold effects, however
a closer examination \cite{vogt} reveals that this is not
true: the density of 
gluons, which controls the heavy quark electroproduction rate, 
is large at small momentum fractions of the proton, 
favoring partonic processes where the initial
gluon has only somewhat more energy than needed to
produce the final state of interest.

As we shall show, for the inclusive cross section and 
for $x$ values larger than about 0.01, 
the higher order QCD corrections 
for this reaction are in fact dominated by singular distributions
$(\alpha_s^n/n! )[\ln^{2n-1}(w)/w]_+$ at order $n$.
The dimensionless weight $w$ is a function of the 
momenta of the external partons, chosen in 
accordance with the kinematics \cite{cls96}, and 
vanishes at threshold. Our main purpose
in this paper is to resum these singular
functions to all orders in perturbation theory
to next-to-leading logarithmic (NLL) accuracy, and, moreover, in
single-particle inclusive (1PI) kinematics.
The required technology has recently been developed in 
Refs.~\cite{cls96,kos1,kos2,los,cmn}.
We also provide analytic results for 
the resummation and finite order 
expansion for pair invariant mass kinematics
(appendix B).
All our analytic results are valid for either charm
or bottom production, but we concentrate our numerical
studies on the case of single inclusive charm production.

The interest in this analysis is first of all intrinsic, 
and lies in studying the quality of the next-to-leading 
logarithmic resummation 
for single-particle inclusive observables.
Second, while the exactly computed $O(\alpha_s)$ corrections 
to the inclusive structure function $F^{\rm charm}_2$ 
\cite{lrsvn93} are, for experimentally accessible $x$ values, 
on average about 30\% - 40\% \cite{hera1}, i.e. non-negligible
but not too alarming, the size of even higher order corrections
bears examination. Our results, in particular our estimates
for the unknown exact next-to-next-to-leading order (NNLO)
$O(\alpha_s^2)$ corrections, 
should help establish the theoretical error on observables 
for the reaction under study, 
and on the gluon density extracted from 
charm quark production at HERA. 
Finally, our results might contribute to a future global 
analysis for resummed parton densities. 

We have organized our paper as follows. Section 2 
contains the derivation of the resummed formula for the 
single-particle inclusive differential cross section for heavy
quark electroproduction. In section 3 we study its finite order 
expansions both analytically and numerically. We examine both 
partonic quantities (coefficient functions), as well as 
hadronic observables, viz. the inclusive structure function $F^{\rm charm}_2$
and its differential distribution in charm transverse momentum.
Section 4 contains our conclusions. 
In appendix A we present some useful formulae related to 
Laplace transformations, and in appendix B we perform
the threshold resummation and NNLO expansion for 
the heavy quark electroproduction cross section in pair-inclusive
kinematics.

\section{Resummed differential cross section}

We begin with the definition of the exact single-particle 
inclusive kinematics. We study the electron ($e$) -- proton ($P$)
reaction
\beq
e(l) + P(p) \rightarrow e(l-q) + {\rm{Q}}(p_1) + X[\overline{\rm{Q}}](p_2')\, ,
\label{epkin}
\eeq
where $X[\overline{\rm{Q}}]$ denotes any allowed hadronic final state
containing at least the heavy antiquark,
${\rm{Q}}(p_1)$ is a heavy quark, 
and $q^\mu$ the momentum transfer of the leptonic to the
hadronic sector of the scattering.
After integrating over the azimuthal angle between
the lepton scattering plane and the tagged heavy
quark plane, neglecting $Z$-boson exchange (we assume
$|q^2| \ll M_Z^2$) and summing over $X$, the cross 
section of the reaction (\ref{epkin}) may be written as \cite{lrsvn93PTdistr}
\beq
\frac{d^4\sigma^{eP\rightarrow eQX}}{dx\,dQ^2\,dT_1\,dU_1} 
= \frac{2\pi\alpha^2}{x\,Q^4}\Bigg[
\Big(1+(1-y)^2\Big) \frac{d^2F_{2,P}(x,Q^2,m^2,T_1,U_1)}{dT_1\,dU_1}
-y^2 \frac{d^2F_{L,P}(x,Q^2,m^2,T_1,U_1)}{dT_1\,dU_1}\Bigg]\,.
\label{crossfs}
\eeq
The functions $d^2F_{k,P}/dT_1\,dU_1,\; k=2,L$ are the double-differential
deep-inelastic heavy quark structure functions and $\alpha$
is the fine structure constant. The kinematic variables $Q^2, x,y$ are 
defined by 
\beq
Q^2 = -q^2 > 0\quad,\quad x=\frac{Q^2}{2p\cdot q}
\quad,\quad y=\frac{p\cdot q}{p\cdot l}\;.
\eeq
We also define the overall invariants
\beq
\label{hadronicinv}
S = (p+q)^2 \equiv S'-Q^2\quad,\quad
T_1 = (p-p_1)^2-m^2 \quad,\quad
U_1 = (q-p_1)^2-m^2\,,
\eeq
and 
\beq
S_4 = S+T_1+U_1+Q^2\,.
\eeq
The structure functions $d^2F_{k,P}/dT_1\,dU_1$ describe the 
strong interaction part of the reaction (\ref{epkin}).
They enjoy the factorization 
\beq
\frac{d^2F_{k,P}(x,S_4,T_1,U_1,Q^2,m^2)}{dT_1\,dU_1}
= {1\over S'^2}\sum_{i=q,\bar{q},g} \,\int\limits_{z^-}^{1}\frac{dz}{z}
\,\phi_{i/P}(z,\mu^2)\;
\omega_{k,i}\Big({x\over z},{s_4\over\mu^2},
                            {t_1\over\mu^2},
                            {u_1\over\mu^2},
                            {Q^2\over\mu^2},
                            {m^2\over\mu^2}, \alpha_s(\mu)
\Big) \;, 
\label{d2ffact}
\eeq
where the sum is over all massless parton flavors, and
$\phi_{i/P}(z,\mu^2)$ is the parton distribution function (PDF)
for flavor $i$ in the proton and $z$ the momentum fraction of parton $i$
in the proton with $z^- = -U_1/(S'+T_1)$. 
The dimensionless functions $\o_{k,i}$ describe the underlying hard parton 
scattering processes and depend on the partonic invariants 
$s^{\prime}, t_1,u_1$ and $s_4$, to be defined in Eq.~(\ref{mandelstam-def1}).
The factorization scale is denoted by $\mu$ and in this paper is always 
taken equal to the renormalization scale.

As we are operating near 
threshold we will assume a fixed number of light flavors
in the evolution of the $\phi_{i/P}$ and the strong coupling
$\alpha_s$.
A further simplification we adopt is neglecting
the contributions from quarks and antiquarks in the sum
over flavors in Eq.~(\ref{d2ffact}). 
This is well-justified as their contribution at NLO was 
found to be a mere 5\% \cite{lrsvn93,hera1}. We therefore only consider 
the gluon-initiated partonic subprocess 
\beql
\g^*(q) +\, g(k) &\longrightarrow& 
{\rm{Q}}(p_1) +\, X'[{\Bar{\rm{Q}}}](p_2')\, ,
\label{photon-gluon-fusion-0}
\eeql
where $k = z\, p$. The partonic invariants are 
\beql
s\,=\,(k+q)^2\,\equiv s'+q^2\, , \hspace*{10mm}t_1\,=\,(k-p_1)^2-m^2\, ,
\hspace*{10mm}u_1\,=\,(q-p_1)^2-m^2\, .
\label{mandelstam-def1}
\eeql
Note that $t_1 = z T_1$, $s' = z(S+Q^2)$ and $u_1=U_1$.
The invariant $s_4 = M_{X'}^2-m^2$ measures the inelasticity of the 
reaction (\ref{photon-gluon-fusion-0}) and is given by
\beql
s'+ t_1 + u_1 &=& s_4~.
\label{mandelstam-sum0}
\eeql
Our goal is to resum those higher order contributions
to  $\omega_{k,g}$ that contain 1PI plus-distributions. 
The latter are defined by
\beq
\left[{\ln^{l}(s_4/m^2)\over s_4}\right]_+
= \lim_{\Delta \rightarrow 0} \Bigg\{
{\ln^{l}(s_4/m^2)\over s_4} \theta(s_4 -\Delta)
+ \frac{1}{l+1}\ln^{l+1}\Big({\Delta\over m^2}\Big)\, \delta({s_4}) \Bigg\}
\label{s4distdef}\,.
\eeq
These distributions are the result of imperfect cancellations
between soft and virtual contributions to the cross section.
At order $O(\alpha^{i+2}_s),\; i=0,1,\ldots$ 
the leading logarithmic (LL) corrections correspond to $l=2i+1$, 
the NLL ones to $l=2i$ etc.

In pursuing our goal, we follow the general principles of Ref.~\cite{cls96},
and the methods of Refs.~\cite{kos1,kos2,los,ks}.
We restrict ourselves from here onwards to the structure
function $F_{2,P}$. The resummation of $F_{L,P}$ demands 
special attention \cite{akhoury}. 
Moreover, $F_{2,P}$ constitutes the bulk of the cross section 
in Eq.~(\ref{crossfs}).

\begin{figure}[hbt]
\begin{center}
\epsfig{file=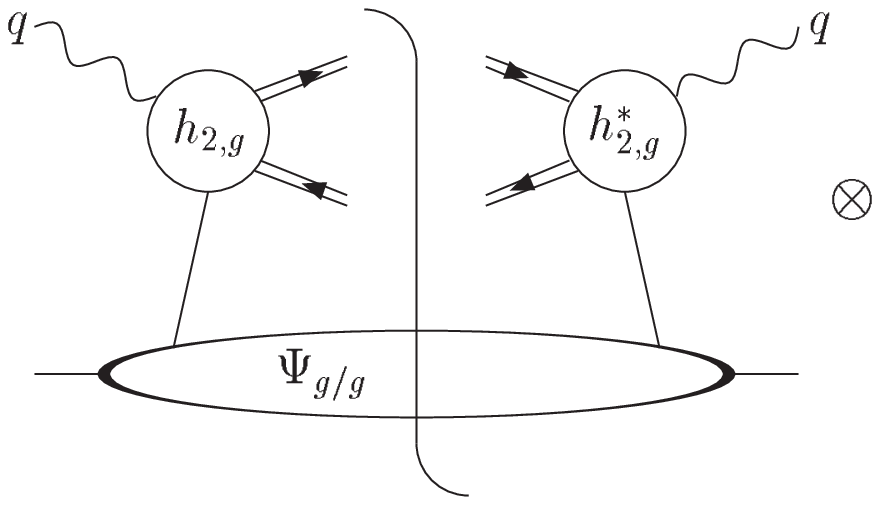,%
bbllx=200pt,bblly=530pt,bburx=435pt,bbury=710pt,angle=0,width=7.0cm}
\epsfig{file=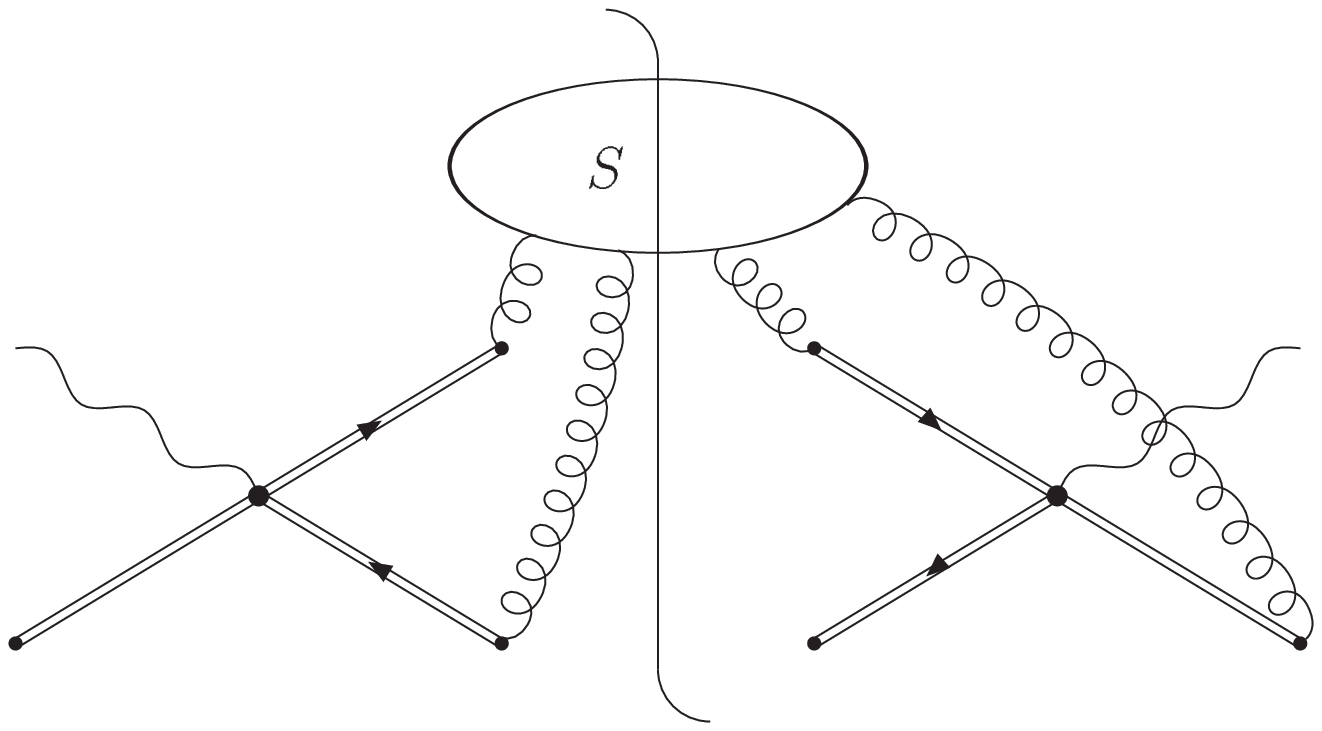,%
bbllx=115pt,bblly=520pt,bburx=490pt,bbury=710pt,angle=0,width=8.0cm}
\caption[dum]{\label{threshconvfig} {\small{
Factorization of heavy quark electroproduction near threshold.
The double lines denote eikonal propagators.}}}
\end{center}
\end{figure}

By replacing the incoming proton in Eq.~(\ref{d2ffact})
with an incoming gluon, one may compute
the hard part $\omega_{2,g}$ in infrared-regulated perturbation
theory.
The resummation of $\omega_{2,g}$ rests upon the
simultaneous factorization of the dynamics and the kinematics of the observable
in the threshold region of phase space. This factorization is pictured in
Fig.~\ref{threshconvfig}. The figure corresponds to the partonic
process (\ref{photon-gluon-fusion-0}) and represents a general partonic 
configuration that produces large corrections,
i.e. corrections containing the singular functions of Eq.~(\ref{s4distdef}).

The figure shows the factorization of the partonic cross section into
various functions, each organizing the large corrections corresponding
to a particular region of phase space. Such factorizations have been discussed
earlier for deep-inelastic scattering, Drell-Yan \cite{sterman87},
heavy quark \cite{ks,bcmn}, dijet \cite{kos1} and for 
general single-particle/jet inclusive cross sections \cite{los}.
The function $\psi_{g/g}$ contains the full dynamics of partons moving
collinearly to the incoming gluon $g$. It includes
all leading and some next-to-leading enhancements.
The function $S(k_S)$ summarizes the dynamics of soft gluons
that are not collinear to $g$, and includes all remaining next-to-leading contributions.
The function $H_{2,g}\equiv h^*_{2,g}h_{2,g}$ incorporates the effects of off-shell partons,
and contains no singular functions. 
There are large next-to-leading corrections associated with the outgoing heavy 
quarks. Were we to treat the heavy quarks as massless, 
each heavy quark would be assigned its own jet function, 
see e.g. Refs.~\cite{kos1,los,cmn}.
In our case the
heavy quark mass prevents collinear singularities, so that all singular
functions arising from the final state heavy quarks are due
to soft gluons. Hence, near threshold, all the singular behavior associated
with the heavy quarks may be included in the soft function \cite{ks}.
We note that $S(k_S)$ is simply a function, and not
a matrix in a space of color tensors, in contrast to 
heavy quark \cite{ks} or jet production \cite{kos1,kos2} 
in hadronic collisions.

The kinematics of reaction (\ref{photon-gluon-fusion-0})
decomposes in a manner that corresponds exactly to the
factorization of Fig.~\ref{threshconvfig}. Momentum conservation
at the parton level means 
\beq
q+ z\,p = p_1 + p_2 + k_S\, ,
\eeq
where, in view of the above discussion, $p_1$ and $p_2$ may be
interpreted as the on-shell momenta of the heavy quark and 
anti-quark respectively.
Squaring and dividing by $m^2$, we have
\beql
{S_4\over m^2} &\simeq& (1-z) \Big({2p\cdot q'-2p\cdot p_1\over m^2}\Big)+ 
{2p_2\cdot k_S \over m^2} \nonumber\\
&\simeq& (1-z) \Big({-u_1\over m^2}\Big) + {s_4\over m^2}
\equiv w_1 \Big({-u_1\over m^2}\Big)+ w_S\, ,
\label{weightrel}
\eeql
where we have dropped terms of order $S_4^2$. 
Notice that the kinematics is in fact 
specified by the dimensionless vector $\zeta^\mu$, defined as
$\zeta^\mu = p_2^\mu/m$. In Eq.~(\ref{weightrel}) we have identified the 
overall single-particle inclusive weight $w$, mentioned in the introduction,
with $S_4/m^2$.
At fixed $S_4$ the infrared-regulated, 
differential partonic structure function factorizes 
\cite{cls96} as
\beql
\label{threshconv}
s'^2 \frac{d^2F_{2,g}(x,S_4,t_1,u_1,Q^2,m^2)}{dt_1\,du_1}
&=&
H_{2,g}(t_1,u_1,Q^2,m^2)\; 
\int dw_1 dw_S\\
&\ & \hspace{-10mm} \times
\delta \left({S_4\over m^2} - w_1 \Big({-u_1\over m^2}\Big)- w_S \right)
\;\psi_{g/g}(w_1,p,\zeta,n)\; S(w_S,\beta_i,\zeta,n)\,, \nonumber
\eeql
where the kinematics relation (\ref{weightrel}) is implemented
in the $\delta$-function.
The $\beta_i$ are the four-velocities of the particles
in the Born approximation to  
Eq.~(\ref{photon-gluon-fusion-0}),
i.e. with $X[\overline{\rm{Q}}](p_2')$ replaced by $\overline{\rm{Q}}(p_2)$.
The various functions are computed in $n\cdot A=0$ gauge.

Further manipulations are carried out most conveniently in terms of Laplace
moments, defined by
\beq
{\tilde f}(N) = \int\limits_0^\infty dw \,e^{-N w} f(w)\,.
\label{laplacetfm}
\eeq
The upper limit of this integral is not so important, and may also be
put at 1, where $\ln w=0$. 

As both the gauge and kinematics vector are timelike,
we may choose \footnote{Note that for lightlike
kinematics, such as in direct photon production, it
may be convenient to keep both vectors separate \cite{los}.} 
$n^\mu=\zeta^\mu$, in which case the 
1PI $\psi_{i/i}$ are equal to the center-of-mass densities 
(for which $\zeta^\mu=n^\mu=\delta^{\mu 0}$) \cite{los}.

Replacing the incoming proton with an incoming gluon in Eq.\ (\ref{d2ffact}) 
and comparing with Eq.~(\ref{threshconv}), we derive
\beql
\tilde{\omega}_{2,g}
\left( N,{t_1\over \mu^2},{u_1\over \mu^2},{Q^2\over\mu^2},{m^2\over\mu^2}\right)
&=&
H_{2,g}\left({t_1\over \mu^2},{u_1\over \mu^2},{Q^2\over\mu^2},{m^2\over\mu^2}\right)\; 
\nonumber\\
&\times&  \left[ {{\tilde\psi}_{g/g}(N\left(-u_1/m^2\right),(p\cdot\zeta/\mu))
\over {\tilde\phi}_{g/g}(N\left(-u_1/m^2\right),\mu)} \right] \;{\tilde S}({m\over N\mu},
\zeta,\beta_i)\; .
\label{threshmoment}
\eeql
Given the factorization in Eqs.~(\ref{threshconv}) and 
(\ref{threshmoment}), the arguments of \cite{cls96,sterman87} ensure that
the $N$-dependence in each of the functions of Eq.\ (\ref{threshmoment})
exponentiates.  

We next discuss each function in Eq.~(\ref{threshconv}), or
(\ref{threshmoment}), in turn, 
starting with the $\j_{g/g}$ wavefunction.
In analogy with the center-of-mass definitions \cite{sterman87,kos1}
it may be defined as an operator matrix element \footnote{For the 
equivalent definition for (anti)quarks, see Ref.~\cite{los}.}
\beql
\j_{g/g}(w,p,\zeta,n) &=& {1\over 2 (N_c^2-1)}
{p\cdot\zeta\over 4\pi (v\cdot p)^2 }
\int\limits_{-\infty}^{\infty} dy\,
e^{-iy\,(1-w)p\cdot\zeta} \nonumber\\
&\times&\langle g(p) | F^{\mu\,\perp}(y\zeta)\;  
\Big[v_{\mu} v_{\nu}\Big]F^{\perp\,\nu}(0)  | g(p)\rangle_{n\cdot A = 0}\, ,
\label{psidef}
\eeql
where $w=1-z$ and $v_\mu$ is a fixed lightlike four-vector in the
opposite direction as $p_\mu$, such that $v\cdot p$
is of order one. The first factor refers to a spin and color average.
The states are normalized such 
that $\langle 0|A^\perp(0)|g(p) \rangle 
= \epsilon^\perp(p)$.
Expression (\ref{psidef}) as a whole is normalized such that 
\beql
\j_{g/g}^{(0)}(w)&=& \d(w)\, .
\eeql
We have calculated the order $\a_s$ corrections to this operator
matrix element in general axial gauge $n\cdot A = 0$.
The $O(\alpha_s)$ 
corrections to the operator matrix element in (\ref{psidef}) are
in $d=4-2\epsilon$ dimensions, and to next-to-leading logarithmic
accuracy
\beql
\j_{g/g}^{(1)}(w,{p\cdot\zeta \over\mu},\epsilon)&=&\frac{\a_s(\mu)}{\p}\,
C_A \Bigg\{ \frac{-1}{\bar\epsilon}\left[\frac{1}{w}\right]_+ + 
\left[\frac{2 \ln(w)}{w}\right]_+ \nonumber \\
&\ &+ \left[\frac{1}{w}\right]_+ \left( \ln\left(\frac{s}{\mu^2}\right) 
- 1 +\ln(2\nu_g)\right) +O(\epsilon) \Bigg\}\, ,
\label{gg-split}
\eeql
where $\nu_g = (\beta_g\cdot n)^2/|n|^2$, with 
$n^\mu$ chosen equal to $\zeta^\mu$, and $\beta^\mu_g$ the
four-velocity of the incoming gluon, defined by
$p^\mu = \beta_g^\mu\sqrt{s/2}$.
Note that $\ln(2\nu_g) + \ln(s/\mu^2) = \ln(4 (p\cdot \zeta)^2/\mu^2)$.
We have abbreviated $1/\bar{\epsilon} = 1/\epsilon-\gamma_E+\ln 4\pi$.
We have verified that the steps for the resummation of the Sudakov double 
logarithms $\ln^2(N)$ in this function trace exactly those for the 
center-of-mass density introduced for the Drell-Yan process in 
Ref.~\cite{sterman87}. 

We choose for $\phi_{g/g}$ the $\overline{\rm MS}$ density,
which does not have Sudakov double logarithms. To the same accuracy as
$\psi_{g/g}$ in Eq.~(\ref{gg-split}), it is given by
\beql
\phi_{g/g} \left(w,\a_s(\mu),\epsilon\right) 
& = & \delta(w)+\frac{\a_s(\mu)}{\p}\,
C_A \left(\frac{-1}{\bar\epsilon}\right)\left[\frac{1}{w}\right]_+ \, .
\label{msbsplit}
\eeql
The resummation of the singular contributions to this function is 
done via the Altarelli-Parisi equation \cite{ap}, see also 
\cite{cls96,KLS}. We only need this 
result in combination with the resummed $\psi_{g/g}$ 
as the ratio $\tilde{\psi}_{g/g}/\tilde{\phi}_{g/g}$ in Eq.~(\ref{threshmoment}).
This ratio is finite, as can readily be verified at one-loop
from Eqs.~(\ref{gg-split}) and (\ref{msbsplit}).

The scale dependence of these functions may be
found as follows \cite{kos1}. The scale dependence of 
$\tilde\psi_{g/g}$ is governed by
\beq
\mu\frac{d}{d\mu}\ln \tilde{\psi}_{g/g}\left(N,\frac{p\cdot\zeta}{\mu},\alpha_s(\mu),\epsilon
\right)
= \gamma_\psi\left(\alpha_s(\mu)\right)\,.
\eeq
The anomalous dimension on the right does not
depend on the moment $N$, as the only
ultraviolet divergences are due to gluon wavefunction renormalization.
Hence, $\gamma_\psi = 2\gamma_g$, where $\gamma_g$ is the 
anomalous dimension of the gluon field in axial gauge.

The anomalous dimension of the $\msb$ density does
depend on $N$, via the Altarelli-Parisi equation
\beq
\mu\frac{d}{d\mu}\ln \tilde{\phi}_{g/g}(N,\a_s(\mu),\epsilon)
= 2 \gamma_{g/g}\left(N,\alpha_s(\mu)\right)\,.
\eeq

We next discuss the soft function ${\tilde S}(N,\zeta)$.
It summarizes the singular contributions arising from soft gluons
that are not collinear to the incoming jet, and is infrared
finite. It depends on the kinematics of the hard
scattering and contributes
only at next-to-leading logarithm.
Threshold resummation for soft functions
is treated in detail in Refs.~\cite{kos2,ks};
a brief sketch will suffice here.
The factorization in Eq.~(\ref{threshconv}) introduces
ultraviolet divergences, distributed in such a way between 
the hard function $H$ and the soft function $S$, that they cancel 
in the product \cite{cls96,kos1}.
The function $S$ may be defined as the matrix element of
a composite operator that connects Wilson lines in the directions
of the external partons. Its extra 
ultraviolet divergences are cancelled by the renormalization of
this operator.
As a result, ${\tilde S}$ obeys the renormalization group equation
\beq
\left(\mu{\partial\over\partial\mu} + \beta(\alpha_s){\partial\over\partial\,\alpha_s}
\right)\,{\tilde S}  = -\left(2{\rm Re}\left\{\Gamma_S(\alpha_s(\mu))\right\}\right){\tilde S}\, ,
\label{sevol}
\eeq
then resums the $\ln(N)$ terms in $\tilde{S}$.
In the reaction under study the anomalous dimension
is a $1\,\times\,1$ matrix in color space,
so we may solve Eq.~(\ref{sevol}) directly:
\beq
{\tilde S}\left({m\over N \mu},\alpha_s(\mu)\right) = {\tilde S}
\left({m\over\m},\alpha_s\left({\m\over N}\right)\right)\;
\exp\left[\,\int\limits_{\mu}^{\m/N}{d\mu'\over\mu'} 2{\rm Re}\left\{\Gamma_S(\alpha_s(\mu'))\right\}
\right]\, .
\label{softfundef}
\eeq
\bigskip
The resummed hard part may now be written
in moment space as 
\beql
\tilde{\omega}_{2,g}
\left( N,{t_1\over \mu^2},{u_1\over \mu^2},{Q^2\over\mu^2},{m^2\over\mu^2}\right)
&=&
H_{2,g}\left({t_1\over \mu^2},{u_1\over \mu^2},{Q^2\over\mu^2},{m^2\over\mu^2}\right)\;
{\tilde S}\left({m\over\m},\alpha_s\left({\m\over N}\right)\right)\;\nonumber\\
&\ & \hspace{-50mm} \times\;
\exp \Bigg \{ E_{(g)}\Big( N \left(\frac{-u_1}{m^2} \right),m^2 \Big) \Bigg \}
\exp\Bigg \{ -2\int\limits_{\mu}^{{m}}{d\mu'\over\mu'}
\left(\gamma_g\left(\alpha_s(\m^{\prime})\right)-
\gamma_{g/g}\Big( N \left(\frac{-u_1}{m^2} \right),\alpha_s(\m^{\prime})\Big)\right)
\Bigg\} \nonumber\\
&\ & \hspace{-50mm}\times\;
\exp\Bigg \{\int\limits_{\mu}^{{{{\mu/N}}}}{d\mu'\over\mu'} 
2\, {\rm Re} \left\{\Gamma_S\left(\alpha_s(\m^{\prime})\right)\right\}\Bigg\}\, .
\label{sigNHSfinal}
\eeql

The first exponent summarizes the $N$-dependence of the
wave function ratio $\tilde{\psi}_{g/g}/\tilde{\phi}_{g/g}$ and 
is, in moment space, the same as for heavy quark and dijet production \cite{kos1,kos2}
\beql
E_{(g)}\left(N_u,m^2\right)
&=&
\int\limits^\infty_0 dw\,\frac{(1-e^{-N_uw})}{w}\; 
\Bigg \{\int\limits^{1}_{w^2} \frac{d\lambda}{\lambda} 
A_{(g)}\left[\alpha_s(\sqrt{\lambda} m)\right]\,\label{omegaexp}
\nonumber\\
&\ &   \hspace{-5mm}
 +\frac{1}{2}\nu_{(g)}\left[\alpha_s(w\,m)\right]\, 
\Bigg \}\, ,\\[2ex]
\nu_{(g)}&=& 2 C_A \frac{\a_s}{\p} 
\Biggl\{ 1 - \ln\left(\frac{4(p \cdot \zeta)^2}{m^2}\right) \Biggr\}\, ,
\label{nudef}
\eeql
where $N_u \equiv N(-u_1/m^2)$, and $A_{(g)}$ is given by the expression
\beq
A_{(g)}(\alpha_s) = C_A\left ( {\alpha_s\over \pi} 
+\frac{1}{2} K \left({\alpha_s\over \pi}\right)^2\right )\, ,
\label{g1def}
\eeq
with $K= C_A\; \left ( {67/ 18}-{\pi^2/ 6 }\right ) - {5/9}n_f$
\cite{kt} and $n_f$ the number of quark flavors.

The second exponent controls the factorization scale dependence of the 
ratio  $\tilde{\psi}_{g/g}/\tilde{\phi}_{g/g}$ through the anomalous dimensions $\gamma_g$ and
$\gamma_{g/g}$. In axial gauge we find them to be
\beql
\gamma_g\left(\alpha_s(\m)\right) &=&  b_2\, \frac{\a_s(\mu)}{\p}\, , \label{psianomdim} \\
\gamma_{g/g}\left(N_u,\alpha_s(\m)\right) &=& - \frac{\a_s(\mu)}{\p} \left( C_A {\rm{ln}}(N_u) - b_2 
\right) +\co(1/N)\,.
\label{MSbaranomdim}
\eeql
where $b_2=(11 C_A - 2n_f)/12$.

\begin{figure}[ht]
\begin{center}
\epsfig{file=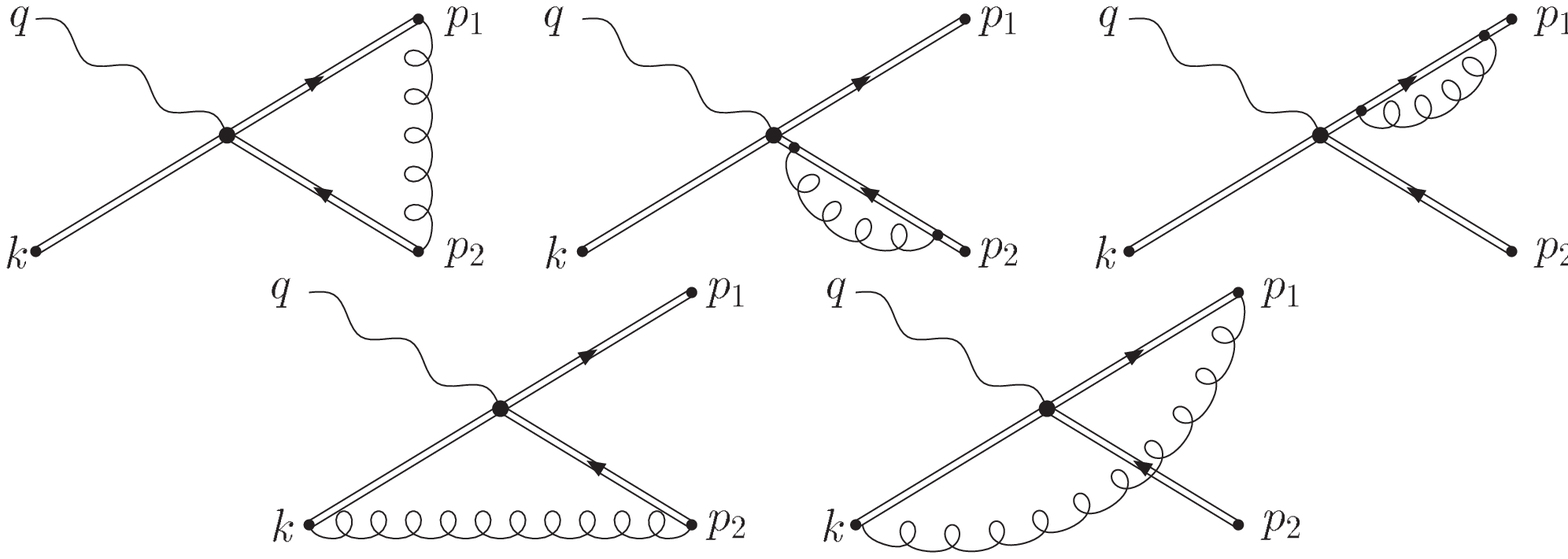,%
bbllx=20pt,bblly=520pt,bburx=585pt,bbury=720pt,angle=0,width=13.0cm}
\caption[dum]{\label{softadim} {\small{
One-loop corrections to the soft function
$S$ for heavy-quark production in photon-gluon fusion.
The double lines denote eikonal propagators. The lines labelled
$k$ corresponds to a gluonic eikonal line, and those labelled
$p_1$ and $p_2$ denote quark and antiquark eikonal lines, respectively.
}}}
\end{center}
\end{figure}

Finally, the $O(\alpha_s)$ expression for the soft anomalous dimension $\Gamma_S$
can be inferred from the UV divergences of the eikonal 
Feynman graphs in Fig.~\ref{softadim}.
We obtain the result
\beql
\G_S&=&\frac{\a_s}{\p}\,
\Biggl\{ C_F \left( - L_\b - 1 \right)  \NO\\[1ex]
& &\hspace*{5mm} -\, \frac{C_A}{2} \left( \ln\left( \frac{4(p \cdot \zeta)^2}{m^2} \right) 
- L_\b - 1 - \ln\left(\frac{-t_1}{m^2}\right) - \ln\left(\frac{-u_1}{m^2}\right) \right) 
\Biggr\}\, .
\label{softad}
\eeql
where $L_\b$ is given by 
\beql
L_\b&=&\frac{1-2\,m^2/s}{\b}
\left\{ \ln\left( \frac{1-\b}{1+\b} \right) + {\rm{i}}\p \right\}\, ,
\;\;\,\,\,\,\,\b=\sqrt{1-4m^2/s}\, ,
\label{Lbetadef}
\eeql
and, for single-particle inclusive kinematics,
\beq
\frac{2p\cdot\zeta}{m} = \frac{-u_1}{m^2}\, .
\eeq

At the conclusion of this section a few remarks are in order.
First, the integral in the exponent in Eq.~(\ref{omegaexp}) can
only be interpreted in a formal sense, as some prescription
should be implemented to avoid integration over the Landau pole
in the running coupling. We do not address such renormalon ambiguities 
here. In this paper 
we only employ the resummed expressions as generating functionals
of approximate perturbation theory.

Second, although our main focus is on 1PI kinematics, 
it is straightforward \cite{los} to derive from 
Eq.~(\ref{sigNHSfinal}) the resummed expression for pair-invariant 
mass kinematics, i.e. the resummed cross section for the process
in Eqs.~(\ref{epkin}) and (\ref{photon-gluon-fusion-0}), differential 
in the ${\rm Q\Bar{Q}}$ invariant mass.
We present this derivation in appendix B.

\section{Finite order results}

In this section we expand our resummed cross section to 
one and two loop order so as to compare with exact NLO calculations 
\cite{lrsvn93,hs} and provide NNLO approximations.

We begin by deriving NLL analytic formulae for the 
single-particle inclusive, {\it partonic} hard part
$\omega_{2,g}$. Subsequently we study their 
behavior in the {\it hadronic} inclusive structure 
function $F_2^{\rm charm}$, and in the 
differential distribution $dF_2^{\rm charm}/dp_T$.

\subsection{Partonic results at NLO and NNLO}

The Born level hard part (see e.g. \cite{gr,lrsvn93}) for 
the process $\g^* + g \longrightarrow
{\rm{Q}} + \Bar{\rm{Q}}$ is 
\beql
\omega_{2,g}^{(0)}(s',t_1,u_1) & = & 
\d(s^{\prime} + t_1 + u_1)\, \s^{\rm Born}_{2,g}(s^{\prime}, t_1, u_1)\, ,
\eeql
with $\s^{\rm Born}_{2,g}$ given by 
\beql
\label{bornfunc1pi}
\s^{\rm Born}_{2,g}(s^{\prime}, t_1, u_1) &=& {\a_s\over 2\p} e_q^2\,
{N_c C_F\over N_c^2-1}\, Q^2 \Biggl[ 
\frac{t_1}{u_1}+\frac{u_1}{t_1}+
4\,\frac{m^2 s^{\prime}}{t_1\, u_1}
\left(1-\frac{m^2 s^{\prime}}{t_1\, u_1}\right) \\
&+&\, 2\, \frac{s^{\prime} q^2}{t_1\, u_1}+2\frac{q^4}{t_1\, u_1}+
2\, \frac{m^2 q^2}{t_1\, u_1}\left(2-\frac{s^{\prime\, 2}}{t_1\, u_1}\right)
+\frac{12 q^2}{s^{\prime}}
\left(\frac{m^2 s^{\prime}}{t_1\, u_1} - 
\frac{s}{s^{\prime}} \right)
\Biggr]\, \NO.
\eeql

We now derive the NLO soft gluons corrections to 
$\omega_{2,g}(s',t_1,u_1)$
by expanding the resummed 
hard part in Eq.~(\ref{sigNHSfinal}) to one loop, using the
explicit expressions for the various functions in Eq.~(\ref{sigNHSfinal}), 
given in Eqs.~(\ref{gg-split}), (\ref{msbsplit}) and 
(\ref{g1def})-(\ref{Lbetadef}). We find
\beql
\omega_{2,g}^{(1)}(s',t_1,u_1) ~\simeq~
K^{(1)}\,\s^{\rm Born}_{2,g}(s^{\prime}, t_1, u_1)\,,
\eeql
where
\beql
\label{oneloopcorr1}
K^{(1)} &=& 
\frac{1}{-u_1}\, 
\left(\j_{\rm{g/g}}^{(1)}(w_1,\e) - \f_{\rm{g/g}}^{(1)}(w_1,\e) \right) 
\Biggr|_{w_1=s_4/(-u_1),\e\rightarrow 0} 
+ \frac{1}{m^2}\left[\frac{1}{w_S}\right]_+\, 2\, {\rm{Re}}\left\{\G_{S}\right\}
\Biggr|_{w_S=s_4/m^2} \\
&=& \frac{\a_s(\m)}{\p}\,
\Biggl[\,
2\, C_A\, \left[{\ln(s_4/m^2)\over s_4}\right]_+ +\,
 \left[{1 \over s_4}\right]_+
\Biggl\{ C_A \left(
\ln\left( \frac{t_1}{u_1} \right) + {\rm{Re}}L_\b 
- \ln\left( \frac{\m^2}{m^2} \right) \right) 
\nonumber \\[1ex]
& &\hspace*{40mm}
- 2\, C_F \left( {\rm{Re}}L_\b + 1 \right)
\Biggr\}
+\, \d(s_4)\,  C_A \ln\left( \frac{-u_1}{m^2} \right)
\ln\left( \frac{\m^2}{m^2} \right)\,\Biggr]\, ,
\label{nloapp}
\eeql
with $\mu$ the $\Bar{{\rm{MS}}}$-mass 
factorization scale and 
$L_\b$ given in Eq.~(\ref{Lbetadef}). The plus-distributions in $s_4$ 
have been defined in Eq.~(\ref{s4distdef}). 
To next-to-leading logarithm, the result in Eq.~(\ref{nloapp}) 
agrees with the exact $\a_s$ corrections of 
Ref.~\cite{lrsvn93}. This holds for the contributions 
both independent and dependent on the factorization scale, 
i.e. terms including powers of $\ln( \m/m)$. 

The scale-dependent logarithms $\ln(\mu/m)$ are explicitly
generated by the second exponent in Eq.~(\ref{sigNHSfinal}), involving
the anomalous dimensions $\gamma_g$ and $\gamma_{g/g}$
given in Eqs.~(\ref{psianomdim}) and (\ref{MSbaranomdim}).
In deriving $\gamma_{g/g}$ in Eq.~(\ref{MSbaranomdim}) we have 
carefully determined all terms which are constant in moment space,
including the $\ln(-u_1/m^2)$ logarithm, absorbed 
in $N_u$. 
In this way we are able to 
determine all factorization scale dependent terms
to NLL accuracy. 

The NNLO corrections may be found in a manner exactly analogous to
what was done for Drell-Yan in Ref.~\cite{magnea} and for 
Higgs production in Ref.~\cite{KLS}.
We expand Eq.~(\ref{sigNHSfinal}) to second order in 
$\a_s$ in moment
space. With the table of Laplace transforms in 
appendix A 
for the singular distributions in Eq.~(\ref{s4distdef}), 
we then find the $\co(\a_s^2)$ threshold-enhanced corrections in 
momentum space. Our result is
\beql
 \omega_{2,g}^{(2)}(s',t_1,u_1)~\simeq~
K^{(2)}\,\s^{\rm Born}_{2,g}(s^{\prime}, t_1, u_1)\,,
\eeql
with
\beql
\label{dsigtwoloop}
\displaystyle
K^{(2)}&=& \frac{\a^2_s(\m)}{\p^2}\,
\Biggl[\,
2\, C_A^2\, \left[{\ln^3(s_4/m^2)\over s_4}\right]_+
\\[1ex]
& &\hspace*{4mm}
\displaystyle
+\, \left[{\ln^2(s_4/m^2)\over s_4}\right]_+
\Biggl\{  3\, C_A^2\,  \left( 
\ln\left( \frac{t_1}{u_1} \right) + {\rm{Re}}L_\b 
- \ln\left( \frac{\m^2}{m^2} \right)
\right) 
\NO\\[1ex]
& &\hspace*{+15mm}
- 2\, C_A 
\left( b_2 + 3\, C_F \left( {\rm{Re}}L_\b + 1 \right) \right)  \Biggr\} 
\NO\\[1ex]
& &\hspace*{4mm}
+\, \left[{\ln(s_4/m^2)\over s_4}\right]_+
\ln\left( \frac{\m^2}{m^2} \right) 
\left\{  C_A^2\, \left( 
- 2 \ln\left( \frac{t_1}{u_1} \right) 
- 2 {\rm{Re}}L_\b + 2\ln\left({-u_1\over m^2}\right) 
+ \ln\left( \frac{\m^2}{m^2} \right)
\right) \right.
\NO\\[1ex]
& &\hspace*{+15mm}
+ 2\, C_A\, 
\left( b_2 + 2\, C_F \left( {\rm{Re}}L_\b + 1 \right) \right)  \Biggr\}
\NO\\[1ex]
& &\hspace*{4mm}
+\, \left[{1 \over s_4}\right]_+
\ln^2\left( \frac{\m^2}{m^2} \right) 
\left\{  - C_A^2 \ln\left({-u_1\over m^2}\right)\, 
 - \frac{1}{2}\, C_A\, b_2 \right\}\, \Biggr]\, .\NO
\eeql
and $b_2=(11 C_A - 2n_f)/12$.

\subsection{Gluon coefficient functions \label{gluoncoefficientfunctions}}

The inclusive structure function $F_{2,P}^{\rm charm}$
is obtained from Eq.~(\ref{d2ffact}) by integrating
over $T_1$ and $U_1$. It can be expanded in 
coefficient functions $c^{(k,l)}_{2,g}$ 
as follows (dropping the $P$ subscript)
\beql
F_2^{\rm charm}(x,Q^2,m^2) \!&=&\!
\frac{\a_s(\m)\, e_{{\rm c}}^2 Q^2}{4 \p^2 m^2}\! 
\int\limits_{ax}^{1}\, dz\, \phi_{g/P}(z,\mu^2)\,
\sum\limits_{k=0}^{\infty} (4 \p \a_s(\m))^k 
\sum\limits_{l=0}^{k} 
c^{(k,l)}_{2,g}(\h,\x) \ln^l\frac{\m^2}{m^2} , \,\,\,\,\,
\label{charmstrucintegrated}
\eeql
where $a=(Q^2+4m^2)/Q^2$ and 
we recall that contributions from light initial state quarks are neglected.
The function $\phi_{g/P}(z,\mu^2)$ denotes the gluon PDF and
the functions $c^{(k,l)}_{2,g}$ depend on the scaling variables
\beql
\h &=& \frac{s}{4 m^2}\, -1\, , \,\,\,\,\,\,  \,\,\,\,\,\, \,\,\,\,\,\,
\x \,=\, \frac{Q^2}{m^2}\, .
\label{etaxidef}
\eeql
The variable $\h$ is a direct measure of the distance to the partonic threshold. 
The inclusive coefficient functions are obtained from Eqs.~(\ref{nloapp})
and (\ref{dsigtwoloop}) by 
\beql
c^{(k,l)}_{2,g}(\h,\x)&=&
\int\limits_{s^{\prime}(1-\b)/2}^{s^{\prime}(1+\b)/2}\,d(-t_1)
\int\limits_{0}^{s_4^{\rm max}}\,ds_4\,\,\,
\frac{d^2 c^{(k,l)}_{2,g}(s^{\prime}, t_1, u_1)}{dt_1\, ds_4}\, ,
\label{partint}
\eeql
where the double differential coefficient functions are in turn related 
to the hard part $\o_{2,g}$ of Eq.~(\ref{d2ffact}) by
\beq
\sum\limits_{l=0}^k 
s'^2\frac{d^2 c^{(k,l)}_{2,g}(s^{\prime}, t_1, u_1)}{dt_1\, ds_4} 
\ln^l\frac{\m^2}{m^2} =
\frac{4\pi^2}{\a_s e_c^2 (4 \p \a_s)^{k}}\, 
\frac{m^2}{Q^2} 
\,\, \omega^{(k)}_{2,g}(s',t_1,u_1) \,
\Biggr|_{u_1=s'+t_1-s_4}\, .
\label{parttodsig}
\eeq
Also, in Eq.~(\ref{partint}) we abbreviated \cite{lrsvn93}
\beql
s_4^{\rm max} &=& \frac{s}{s^{\prime}\, t_1}\, 
\left(t_1 + \frac{s^{\prime}(1-\b)}{2} \right)
\left(t_1 + \frac{s^{\prime}(1+\b)}{2} \right)\, .
\eeql

\bigskip
\bigskip

We begin our numerical studies with the 
gluon coefficient functions $c^{(k,0)}_{2,g}(\h,\x)$,
i.e. those that are not accompanied by scale-dependent logarithms.
We recall that all our results are derived in the $\msb$ scheme.
The functions $c^{(0,0)}_{2,g}$ and  $c^{(1,0)}_{2,g}$ are known
exactly, see Ref.~\cite{lrsvn93}. The one-loop expansion of our 
resummed hard part provides an approximation to the
exactly known $c^{(1,0)}_{2,g}$. To judge the 
benefits of resumming to {\it next-to-}leading logarithmic 
as compared to leading logarithmic accuracy, we can distill 
also LL expressions, by keeping only
the $[\ln(s_4/m^2)/s_4]_+$ and $[\ln^3(s_4/m^2)/s_4]_+$ terms
in Eqs.~(\ref{nloapp}) and (\ref{dsigtwoloop})
respectively.

\begin{figure}
\begin{center}
\epsfig{file=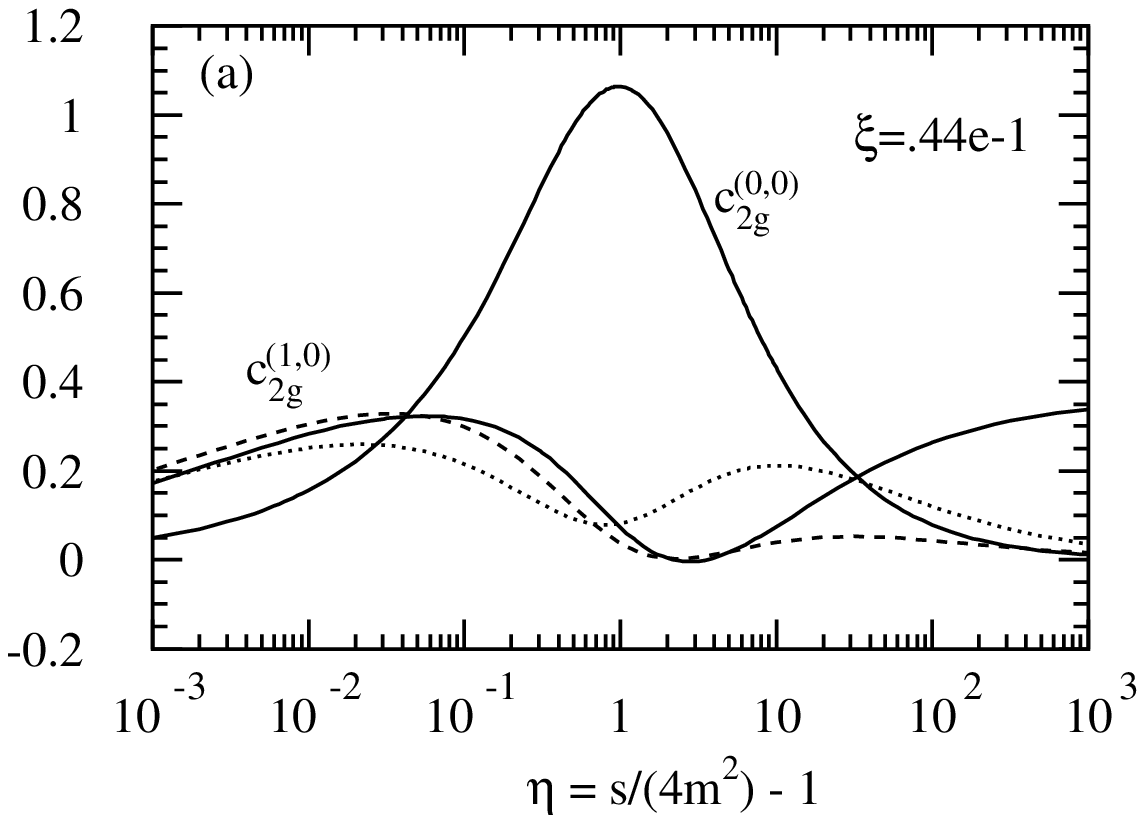,%
bbllx=50pt,bblly=130pt,bburx=285pt,bbury=470pt,angle=270,width=8.25cm}
\epsfig{file=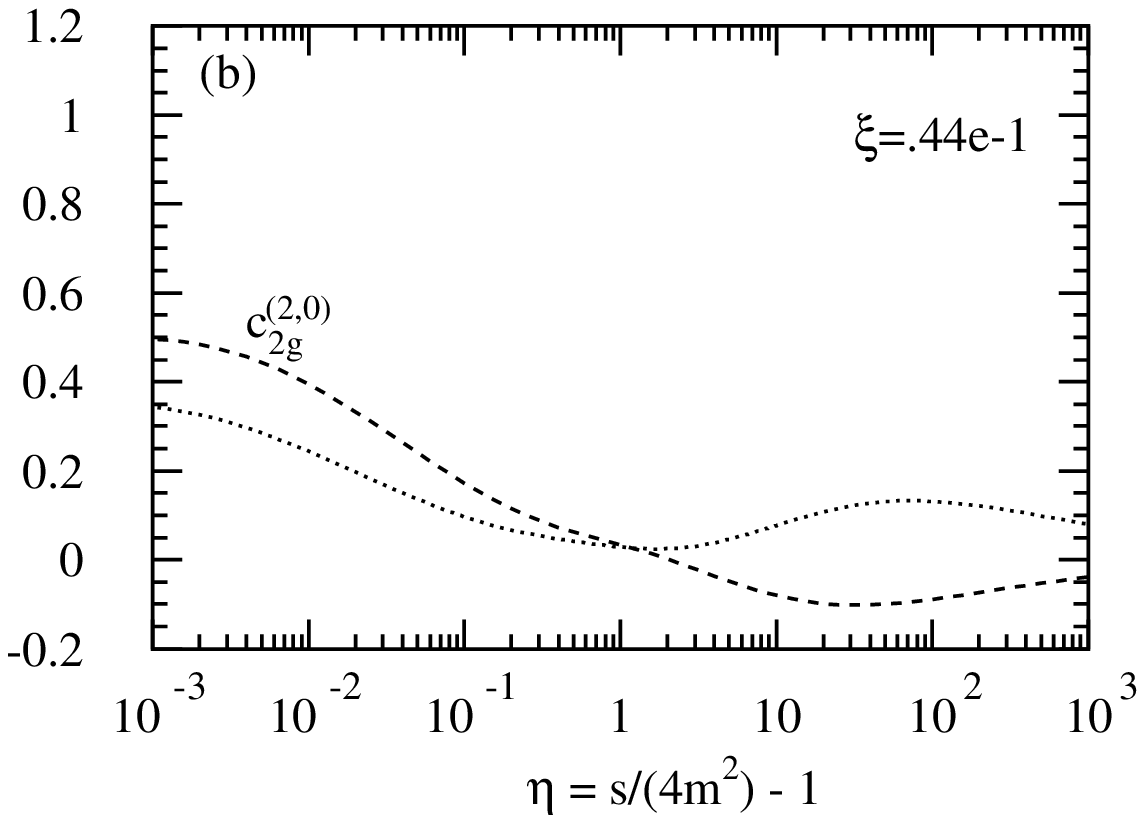,%
bbllx=50pt,bblly=110pt,bburx=285pt,bbury=450pt,angle=270,width=8.25cm}
\caption[dum]{\label{plot-onem1} {\small{
(a): The $\h$-dependence of the coefficient functions 
$c^{(k,0)}_{2,g}(\h,\x),\;k=0,1$   
for $Q^2=0.1\,{\rm GeV}^2$ with $m=1.5\,{\rm GeV}$. 
Plotted are the exact results for $c^{(k,0)}_{2,g},\;k=0,1$ 
(solid lines), the LL approximation to $c^{(1,0)}_{2,g}$ (dotted line) 
and the NLL approximation to $c^{(1,0)}_{2,g}$ (dashed line). 
(b): The $\h$-dependence of the coefficient function 
$c^{(2,0)}_{2,g}(\h,\x)$ for $Q^2=0.1\,{\rm GeV}^2$ with $m=1.5\,{\rm GeV}$. 
Plotted are the LL approximation (dotted line) 
and the NLL approximation (dashed line).}}}
\end{center}
\end{figure}
\begin{figure}
\begin{center}
\epsfig{file=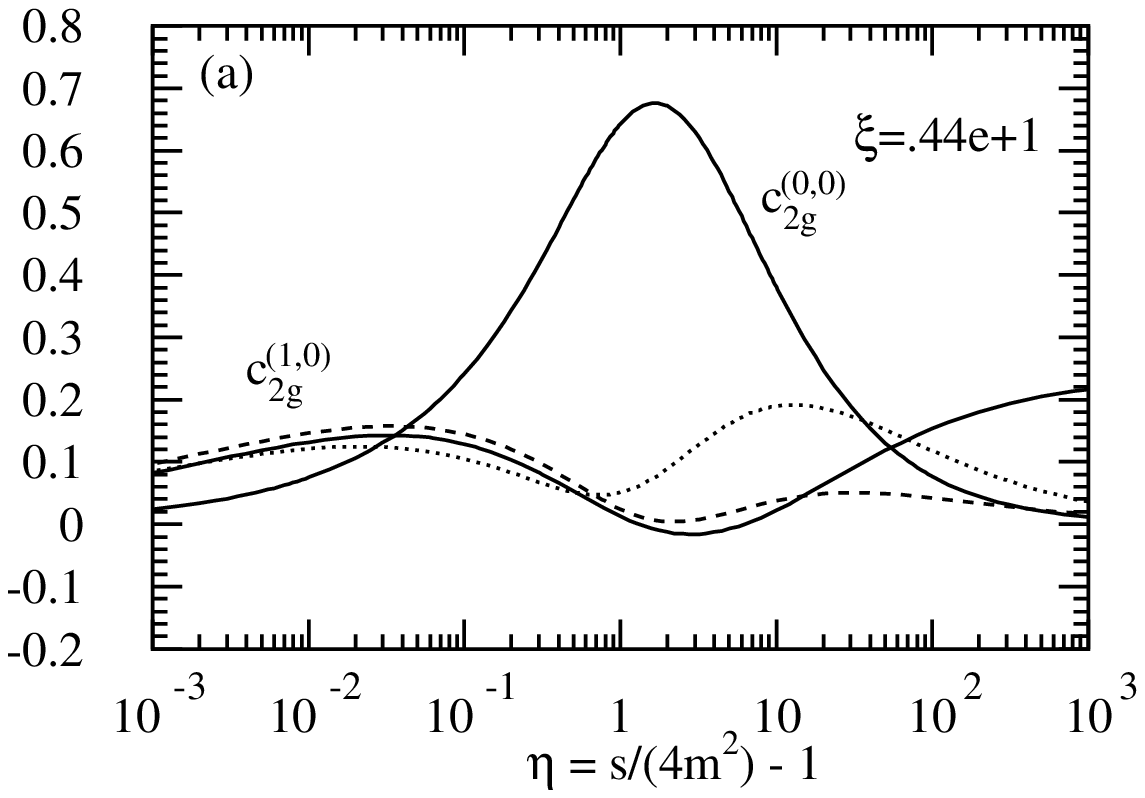,%
bbllx=50pt,bblly=130pt,bburx=285pt,bbury=470pt,angle=270,width=8.25cm}
\epsfig{file=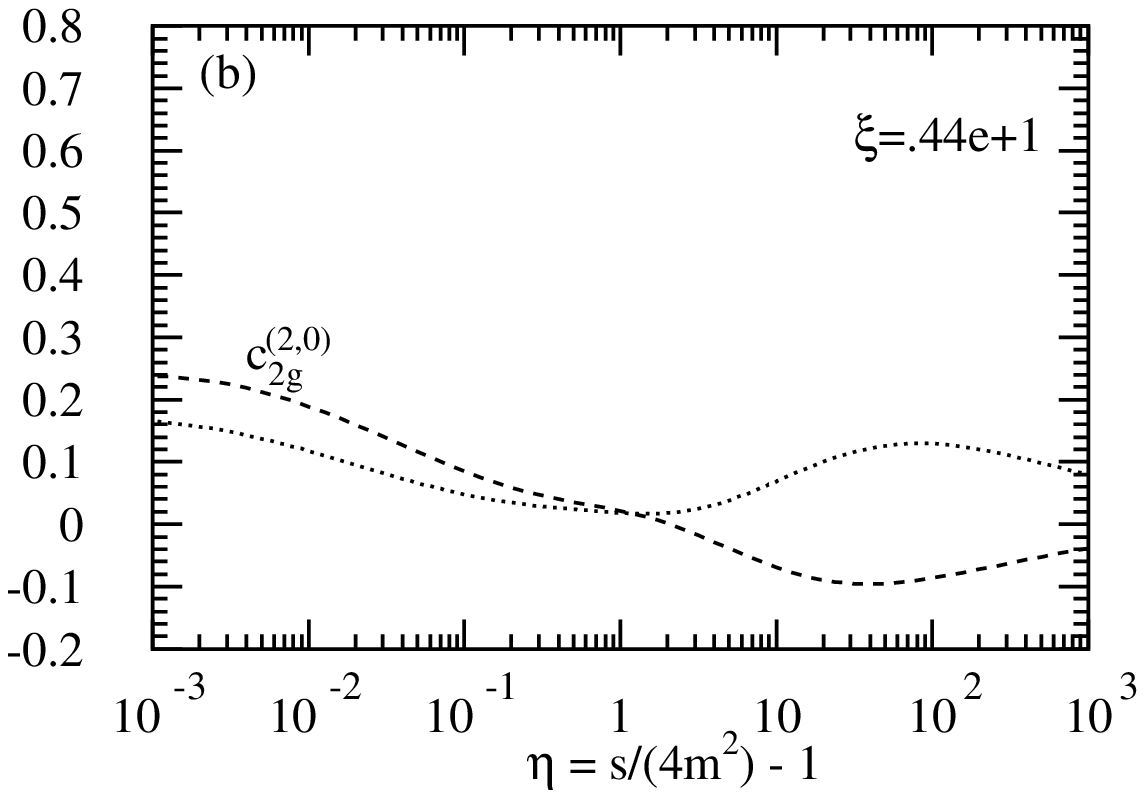,%
bbllx=50pt,bblly=110pt,bburx=285pt,bbury=450pt,angle=270,width=8.25cm}
\caption[dum]{\label{plot-onep1} {\small{
(a): The $\h$-dependence of the coefficient functions 
$c^{(k,0)}_{2,g}(\h,\x),\;k=0,1$   
for $Q^2=10\,{\rm GeV}^2$ with $m=1.5\,{\rm GeV}$. 
The notation is the same as in Fig.~\ref{plot-onem1}a. 
(b): The $\h$-dependence of the coefficient function 
$c^{(2,0)}_{2,g}(\h,\x)$ for $Q^2=10\,{\rm GeV}^2$ with $m=1.5\,{\rm GeV}$. 
The notation is the same as in Fig.~\ref{plot-onem1}b.}}}
\end{center}
\end{figure}
\begin{figure}
\begin{center}
\epsfig{file=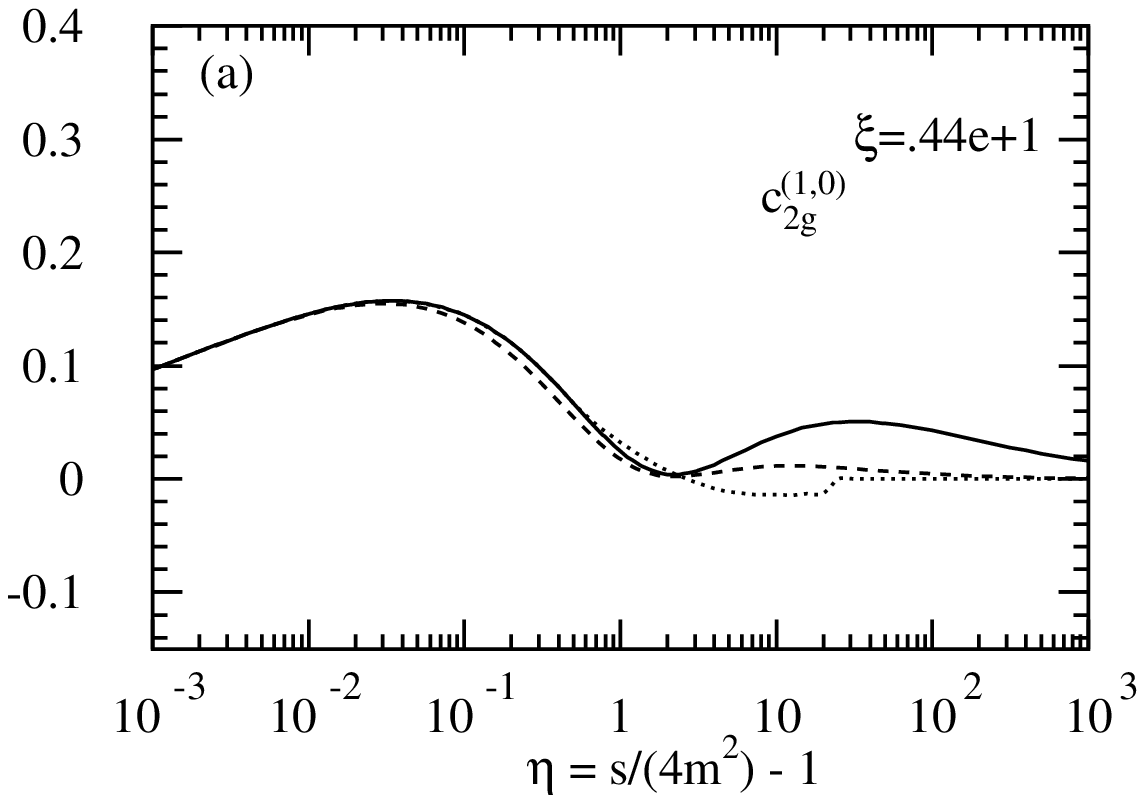,%
bbllx=50pt,bblly=130pt,bburx=285pt,bbury=470pt,angle=270,width=8.25cm}
\epsfig{file=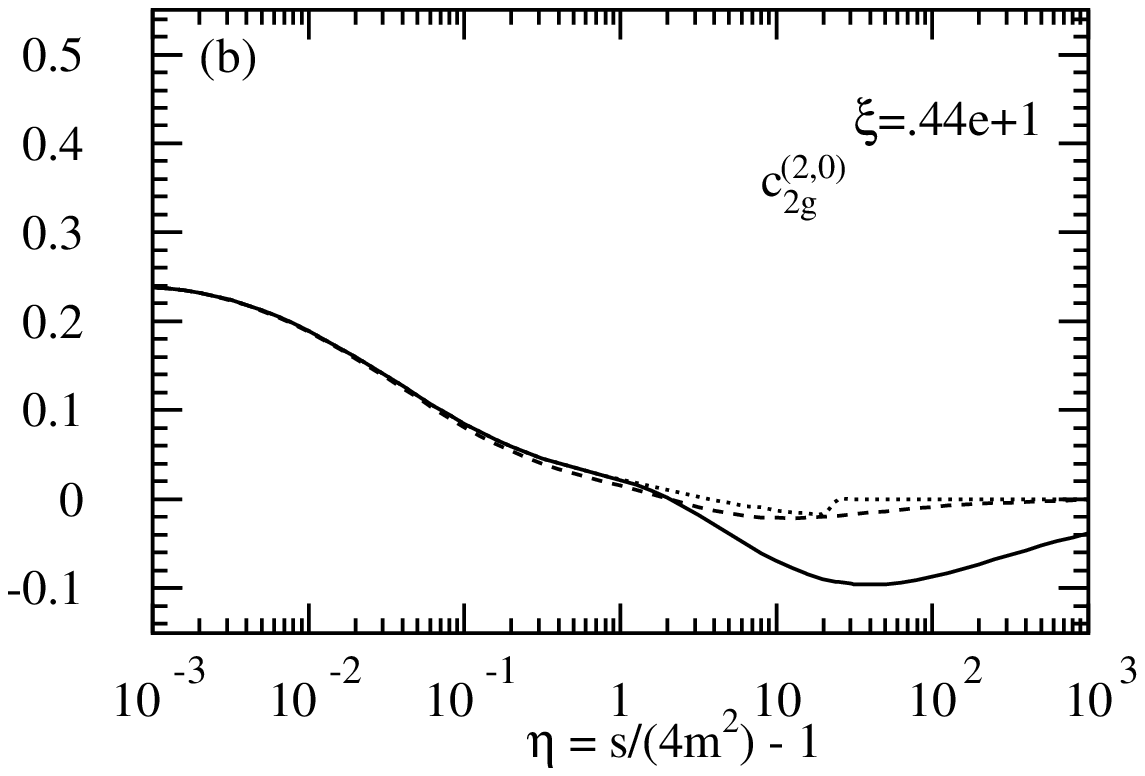,%
bbllx=50pt,bblly=110pt,bburx=285pt,bbury=450pt,angle=270,width=8.25cm}
\caption[dum]{\label{plot-onep1veto} {\small{
(a): The NLL approximation to the 
coefficient function  $c^{(1,0)}_{2,g}(\h,\x)$   
for $Q^2=10\,{\rm GeV}^2$ and $m=1.5\,{\rm GeV}$ 
with restrictions to the small $\h$-region. 
Plotted are the unmodified NLL result (solid line), 
the NLL result with a factor $\theta(m^2-s_4)$ included (dotted line) 
and the NLL result multiplied with a damping factor $1/\sqrt{1+\eta}$ 
(dashed line). 
(b): The $\h$-dependence of the NLL approximation to the 
coefficient function  $c^{(2,0)}_{2,g}(\h,\x)$   
for $Q^2=10\,{\rm GeV}^2$ and $m=1.5\,{\rm GeV}$. 
The notation is the same as in Fig.~\ref{plot-onep1veto}a.}}}
\end{center}
\end{figure}

In Figs.~\ref{plot-onem1}a and \ref{plot-onep1}a we plot 
the functions $c^{(k,0)}_{2,g}(\h,\x),\;k=0,1$ versus
$\h$, for two values of $\x=Q^2/m^2$. 
For the coefficient functions only
the ratio $\xi$ matters, however, we chose those values to correspond 
to $Q^2=0.1$ and  $10$ ${\rm GeV}^2$ for a charm
mass \footnote{In the next subsection, 
where we study the hadronic structure function, we use
the value $m=1.6\,{\rm GeV}$.}
of $m=1.5\,{\rm GeV}$.
These figures reveal that, at one loop, the LL accuracy provides 
a good approximation for very small $\h$, however, with significant 
deviations from the exact result for larger $\h$, 
i.e. as the distance from threshold increases.
On the other hand, the NLL approximation is excellent 
over a much wider range in $\h$, 
up to values of about 10. 

We also show $c^{(2,0)}_{2,g}$, obtained from 
Eqs.~(\ref{dsigtwoloop}) and (\ref{parttodsig}), in LL and NLL
approximation in Figs.~\ref{plot-onem1}b and \ref{plot-onep1}b  
for the same values of $m$ and $Q^2$ as before. 
We observe more structure than in the 
$c^{(1,0)}_{2,g}$ curves. 
Figs.~\ref{plot-onem1}b and \ref{plot-onep1}b 
show some sizable deviations from zero at large $\h$,
which leads us to consider the following issue. 
Kinematically, larger $\h$ values allow 
$w=s_4/m^2 \gg 1$, whereas, in contrast, 
the threshold region is defined by $w\approx 0$. 
Therefore we should consider the inclusion of a 
factor $\theta(m^2-s_4)$ in the integral (\ref{partint}) in order to 
suppress spurious large corrections in the large $\h$ region.

In Figs.~{\ref{plot-onep1veto}a and {\ref{plot-onep1veto}b 
we show the effect of the veto $\theta(m^2-s_4)$ 
on the NLL coefficient functions $c^{(k,0)}_{2,g}$ for $k=1,2$ 
for values of $Q^2=10\,{\rm GeV}^2$ and $m=1.5\,{\rm GeV}$.
Also shown is the effect of a damping factor $1/\sqrt{1+\eta}$
\cite{mengetal}, which can be included
outside the integral in Eq.~(\ref{partint}) and is therefore easier
to implement. We see that the effect of the veto is
well-mimicked by the damping factor, therefore we shall use the
latter, rather than the veto, where needed in what follows. 
Of course, for smaller values of $\h$ the damping factor leads to
a slight underestimation of the coefficient functions compared 
to the veto, but these effects are negligible,
as Fig.~{\ref{plot-onep1veto} shows.

\bigskip
\bigskip

\begin{figure}
\begin{center}
\epsfig{file=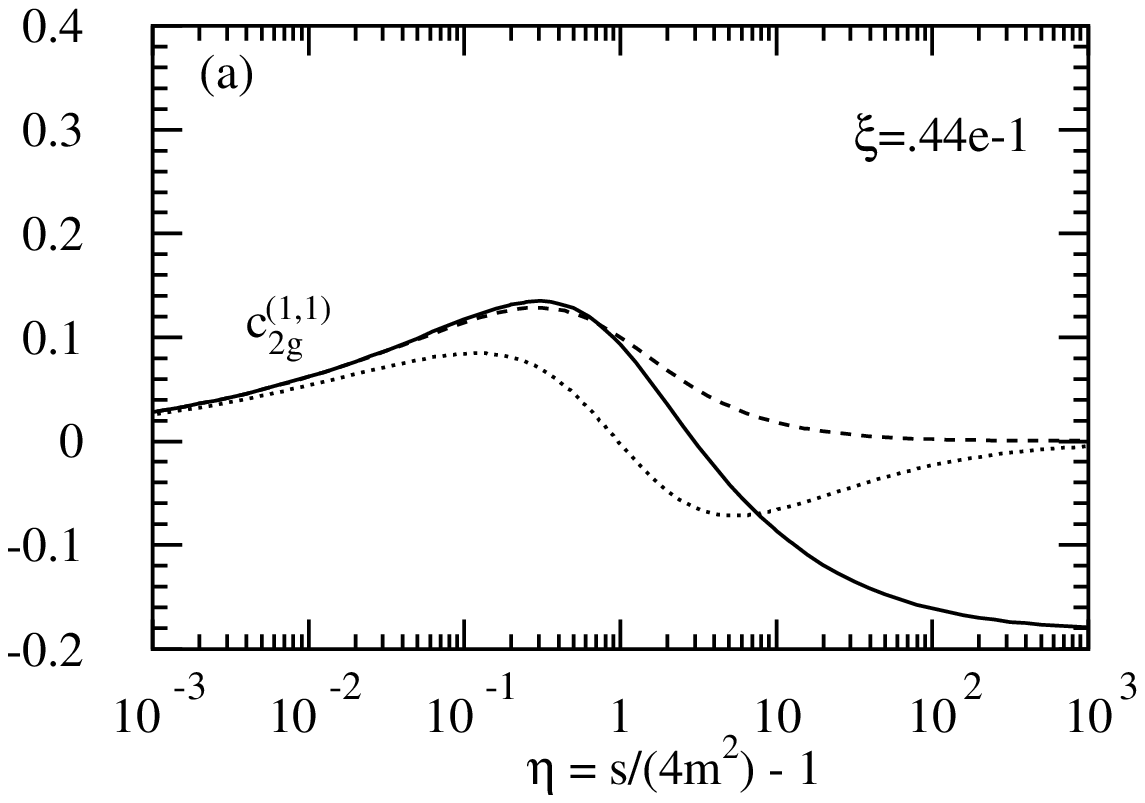,%
bbllx=50pt,bblly=130pt,bburx=285pt,bbury=470pt,angle=270,width=8.25cm}
\epsfig{file=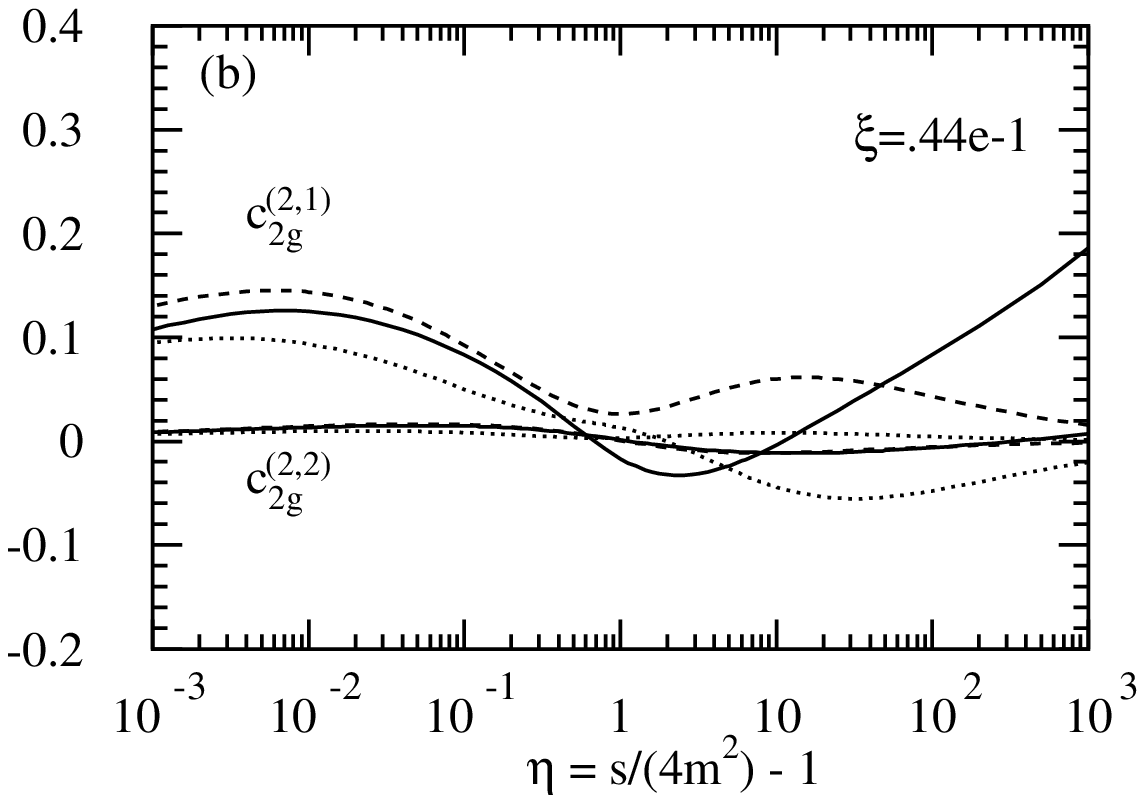,%
bbllx=50pt,bblly=110pt,bburx=285pt,bbury=450pt,angle=270,width=8.25cm}
\caption[dum]{\label{plot-twom1} {\small{
(a): The $\h$-dependence of the coefficient function 
$c^{(1,1)}_{2,g}(\h,\x)$   
for $Q^2=0.1\,{\rm GeV}^2$ with $m=1.5\,{\rm GeV}$. 
Plotted are the exact result (solid line), 
the LL approximation (dotted line) 
and the NLL approximation (dashed line). 
(b): The $\h$-dependence of the coefficient functions 
$c^{(2,l)}_{2,g}(\h,\x),\;l=1,2$ for $Q^2=0.1\,{\rm GeV}^2$ 
with $m=1.5\,{\rm GeV}$. 
Plotted are the exact results (solid lines), 
the LL approximations (dotted lines) 
and the NLL approximations (dashed lines).}}}
\end{center}
\end{figure}
\begin{figure}
\begin{center}
\epsfig{file=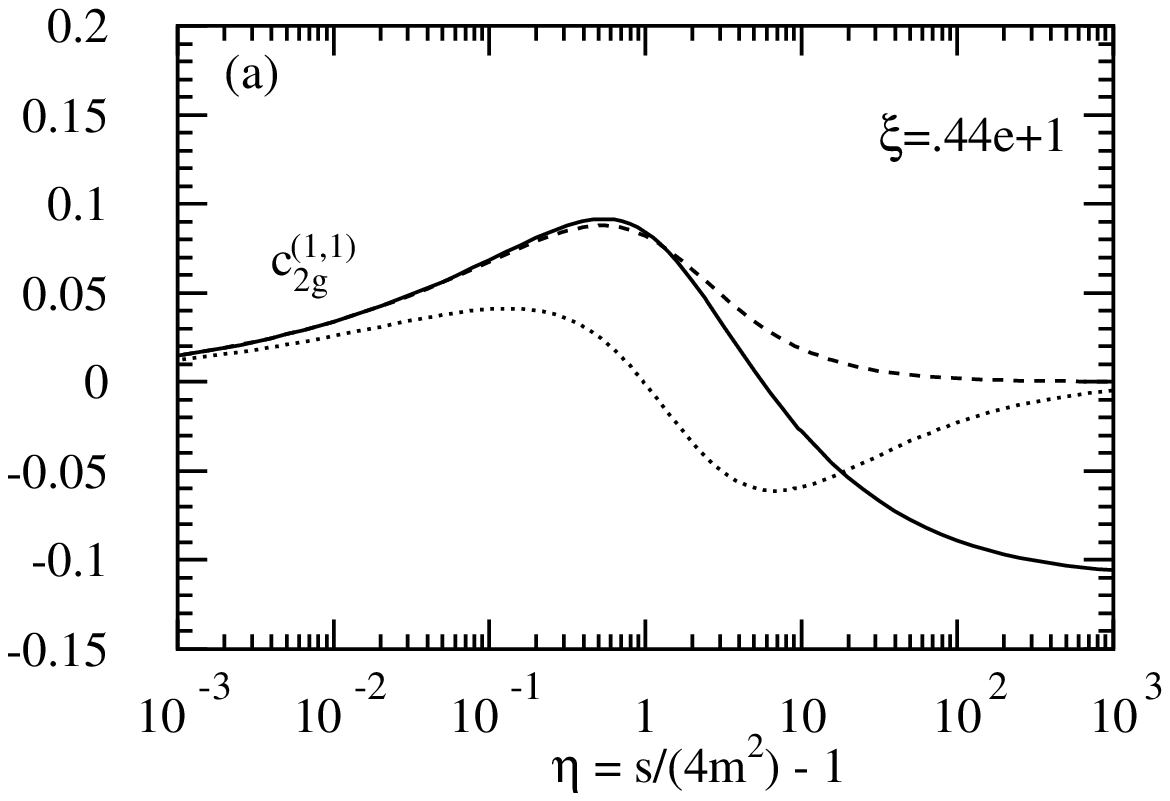,%
bbllx=50pt,bblly=130pt,bburx=285pt,bbury=470pt,angle=270,width=8.25cm}
\epsfig{file=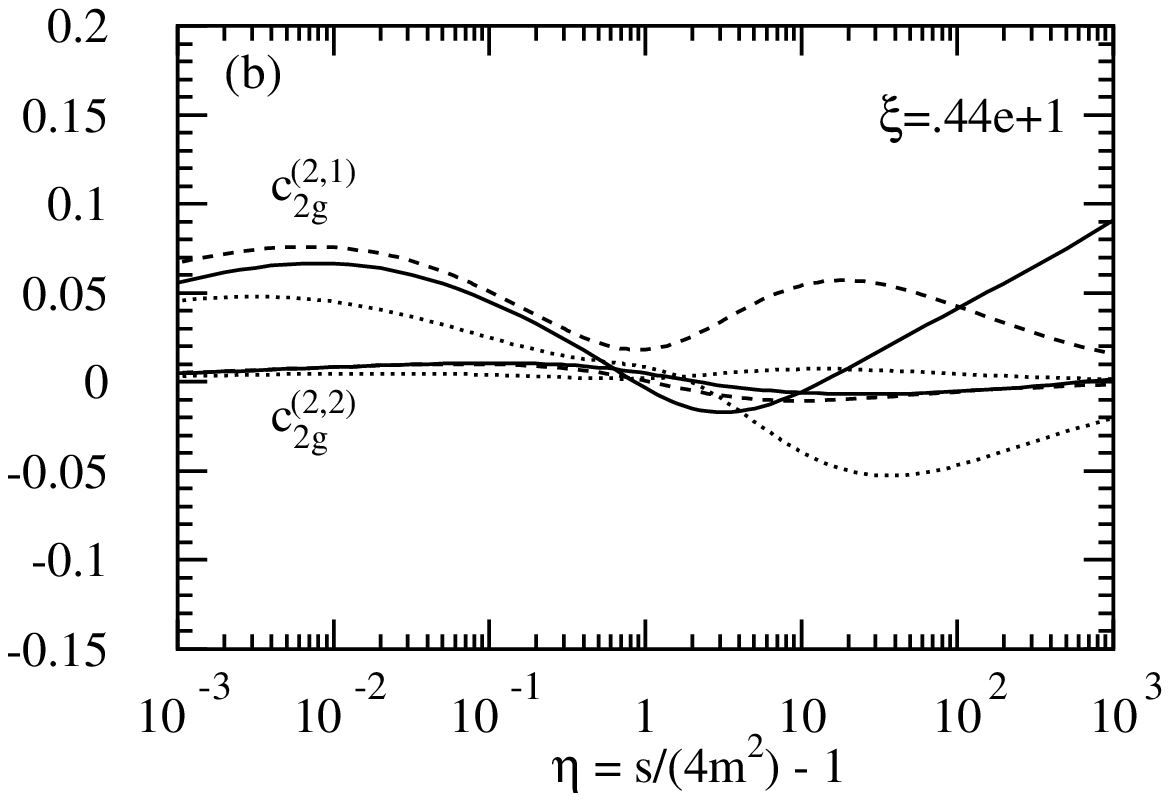,%
bbllx=50pt,bblly=110pt,bburx=285pt,bbury=450pt,angle=270,width=8.25cm}
\caption[dum]{\label{plot-twop1} {\small{
(a): The $\h$-dependence of the coefficient function 
$c^{(1,1)}_{2,g}(\h,\x)$   
for $Q^2=10\,{\rm GeV}^2$ with $m=1.5\,{\rm GeV}$. 
The notation is the same as in Fig.~\ref{plot-twom1}a. 
(b): The $\h$-dependence of the coefficient functions 
$c^{(2,l)}_{2,g}(\h,\x),\;l=1,2$ for $Q^2=0.1\,{\rm GeV}^2$ 
with $m=1.5\,{\rm GeV}$. 
The notation is the same as in Fig.~\ref{plot-twom1}b.}}}
\end{center}
\end{figure}
\begin{figure}
\begin{center}
\epsfig{file=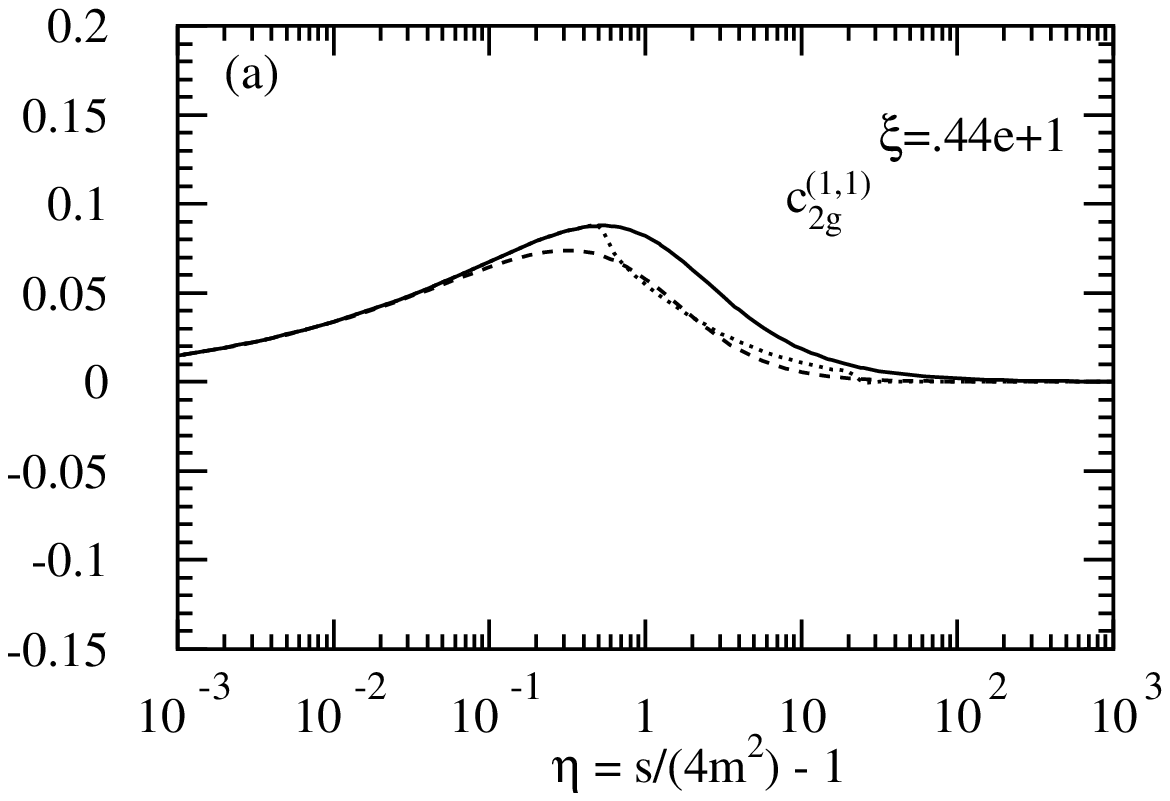,%
bbllx=50pt,bblly=130pt,bburx=285pt,bbury=470pt,angle=270,width=8.25cm}
\epsfig{file=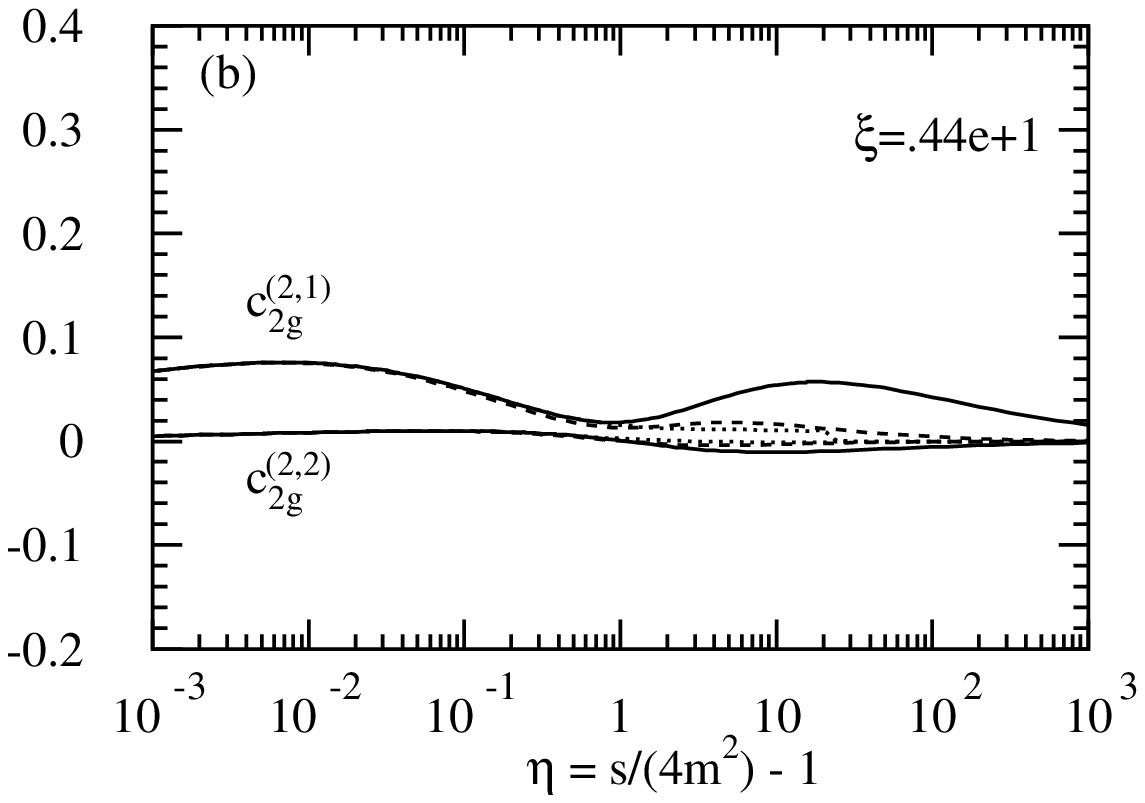,%
bbllx=50pt,bblly=110pt,bburx=285pt,bbury=450pt,angle=270,width=8.25cm}
\caption[dum]{\label{plot-twop1veto} {\small{
(a):  The NLL approximation to the 
coefficient function  $c^{(1,1)}_{2,g}(\h,\x)$   
for $Q^2=10\,{\rm GeV}^2$ and $m=1.5\,{\rm GeV}$ 
with restrictions to the small $\h$-region. 
The notation is the same as in Fig.~\ref{plot-onep1veto}a
(b): The NLL approximation to the 
coefficient function  $c^{(2,l)}_{2,g}(\h,\x),\, l=1,2$ 
for $Q^2=10\,{\rm GeV}^2$ and $m=1.5\,{\rm GeV}$ 
with restrictions to the small $\h$-region. 
The notation is the same as in Fig.~\ref{plot-onep1veto}a.}}}
\end{center}
\end{figure}

Let us now investigate those coefficient functions 
$c^{(k,l)}_{2,g}(\h,\x),\, l\geq 1$ in Eq.~(\ref{charmstrucintegrated}) 
that determine the dependence of $F_2^{\rm charm}$ 
on the mass factorization scale $\m$. 
As mentioned earlier, this is of some relevance since the exact 
NLO $F_2^{\rm charm}$ exhibits considerably more $\mu$ sensitivity 
at large $x$ (near threshold), than at small $x$, cf. Ref.~\cite{vogt}. 
The natural question arises whether and, if so how much, 
our approximate NNLO results might improve the situation.

Explicitly, the coefficient functions under 
consideration up to NNLO are $c^{(1,1)}_{2,g}$ 
which is known exactly \cite{lrsvn93} 
and the previously unknown functions $c^{(2,1)}_{2,g}$ and $c^{(2,2)}_{2,g}$. 
We can infer the exact results for 
these functions from renormalization group 
arguments. We find {\footnote{Eq.~(\ref{ex-c11}) 
agrees with the result for $c^{(1,1)}_{2,g}$ in Ref.~\cite{lrsvn93}.}}
\beql
c^{(1,1)}_{2,g} &=& \frac{1}{4 \p^2}
\left[ b_2\, c^{(0,0)}_{2,g} - \frac{1}{2} 
c^{(0,0)}_{2,g} \otimes  P_{gg}^{(0)} \right]\, ,
\label{ex-c11}\\[1ex]
c^{(2,1)}_{2,g} &=& 
\frac{1}{(4 \p^2)^2}
\left[ b_3\, c^{(0,0)}_{2,g} - \frac{1}{2} 
c^{(0,0)}_{2,g} \otimes  P_{gg}^{(1)} \right]\, 
+ \frac{1}{4 \p^2}
\left[  2 b_2\, c^{(1,0)}_{2,g} - \frac{1}{2} 
c^{(1,0)}_{2,g} \otimes P_{gg}^{(0)} \right]\, ,
\label{ex-c21}\\[1ex]
c^{(2,2)}_{2,g} &=& \frac{1}{(4 \p^2)^2}
\left[ b_2^2\, c^{(0,0)}_{2,g} 
- \frac{3}{4} b_2\,  c^{(0,0)}_{2,g} \otimes P_{gg}^{(0)} + 
\frac{1}{8} c^{(0,0)}_{2,g} \otimes P_{gg}^{(0,0)} \right]\, ,
\label{ex-c22}
\eeql
with $\b$-function coefficients $b_2=(11 C_A - 2n_f)/12$, 
$b_3=(34 C_A^2- 6 C_F n_f - 10 C_A n_f)/48$ and 
the one- and two-loop 
gluon-gluon splitting functions $P_{gg}^{(0)}$, $P_{gg}^{(1)}$, 
cf. Ref.~\cite{fp80}. 
The convolutions involving a coefficient function $c^{(i,0)}_{2,g}$
are defined as 
\beql
\left( c^{(i,0)}_{2,g} \otimes P_{gg}^{(j)} \right) (\h(x),\x)
&\equiv& \int\limits_{ax}^{1}\, dz\,\, 
c^{(i,0)}_{2,g} \left(\h\left(\frac{x}{z}\right),\x\right)\,  
P_{gg}^{(j)}(z)\,  , 
\eeql
where $a=(Q^2+4m^2)/Q^2$ 
and $\h(x)$ is derived from Eq.~(\ref{etaxidef}), 
$\h(x) = \x/4 (1/x-1) - 1$. 
Finally, the function $P_{gg}^{(0,0)}$ in Eq.~(\ref{ex-c22}) is given by 
the standard convolution of two splitting functions $P_{gg}^{(0)}$, 
\beql
P_{gg}^{(0,0)}(x) 
&\equiv& 
\int\limits_{0}^{1}\, dx_1\,
\int\limits_{0}^{1}\, dx_2\, \d(x -x_1 x_2)
P_{gg}^{(0)}(x_1) P_{gg}^{(0)}(x_2)\, .
\eeql

With Eqs.~(\ref{ex-c11})-(\ref{ex-c22}) in hand, 
we are able to check the quality of our NLL approximation 
at NLO {\it and} NNLO against exact answers. 
Again, to fully appreciate the improvement of NLL resummation, 
we truncate our approximation to LL accuracy, keeping only
the $\ln(\m/m) [1/s_4]_+$ term in Eq.~(\ref{nloapp}) and the terms  
$\ln(\m/m) [\ln^2(s_4/m^2)/s_4]_+$ and $\ln^2(\m/m) [\ln(s_4/m^2)/s_4]_+$ 
in Eq.~(\ref{dsigtwoloop}), respectively.

In Figs.~\ref{plot-twom1}a and \ref{plot-twop1}a we plot 
the function $c^{(1,1)}_{2,g}(\h,\x)$ versus 
$\h$, for two values of $\x=Q^2/m^2$,
while Figs.~\ref{plot-twom1}b and \ref{plot-twop1}b display 
the functions $c^{(2,l)}_{2,g},\;l=1,2$. 
The evaluation of Eq.~(\ref{ex-c21}) has been 
performed with the help of the parametrization for 
$c^{(1,0)}_{2,g}$ from Ref.~\cite{rsvn95}.
{}From Figs.~\ref{plot-twom1} and \ref{plot-twop1}, we see again
that the NLL approximations are superior to the 
LL ones. In Figs.~\ref{plot-twop1veto}a and 
\ref{plot-twop1veto}b we show the effect of the veto 
$\theta(m^2-s_4)$ in the integral (\ref{partint}) and the 
damping factor $1/\sqrt{1+\eta}$ on $c^{(1,1)}_{2,g}(\h,\x)$ and 
$c^{(2,l)}_{2,g}(\h,\x),\, l = 1,2$ 
for values of $\xi$ corresponding to $Q^2=10\,{\rm GeV}^2$ and 
$m=1.5\,{\rm GeV}$. 
In the following sections we shall for the sake of uniformity 
include a damping factor for the $c^{(2,l)}_{2,g}(\h,\x),\, l = 1,2$
{\it and} $c^{(1,1)}_{2,g}(\h,\x)$, even though the undamped NLL
$c^{(1,1)}_{2,g}$ approximates the exact $c^{(1,1)}_{2,g}$
somewhat better, as a comparison of Figs.~\ref{plot-twop1}a and 
\ref{plot-twop1veto}a shows. We verified that the difference
is negligible at the hadronic level.

\subsection{Inclusive structure function
$F_2^{\rm charm}$ \label{charmstructurefunction}}

Having obtained encouraging results for the quality of the partonic
approximations Eqs.~(\ref{nloapp}) and (\ref{dsigtwoloop}) 
at the inclusive level, we shall now assess their effect 
at the hadronic level. 

Throughout we use the CTEQ4M \cite{cteq4} gluon density. For NLO plots 
we use the two-loop expression for $\a_s$ with $n_f = 4$, $\L = 0.298\,{\rm GeV}$. 
Respectively, for NNLO plots, the three-loop expression \cite{threeloopalphas}
for $\a_s$ and an adjusted $\L = 0.265\,{\rm GeV}$.

In Fig.~\ref{plot-pdf-1}a and \ref{plot-pdf-1}b we plot the gluon density
$\f_{g/P}(x/z,\m^2)$ as a function of $\h$ on the same scale
as the coefficient functions $c^{(k,l)}_{2,g}(\h,\x)$ for 
$x=0.1$ and $0.01$ and three different choices of the factorization scale 
($\m = m, 2m, \sqrt{Q^2 + 4 m^2}$).
From now on we use a charm mass of $m =1.6\,{\rm GeV}$ 
and take $Q^2 = 10\,{\rm GeV}$. Our heavy quark mass is always 
a pole mass, as in Ref.~\cite{lrsvn93}.
The figure shows that the gluon density indeed provides support for 
the coefficient functions in Eq.~(\ref{charmstrucintegrated}) 
in the threshold region for $x\geq 0.01$, 
although the figure also shows the extent of the 
support to be somewhat scale sensitive. 

\begin{figure}
\begin{center}
\epsfig{file=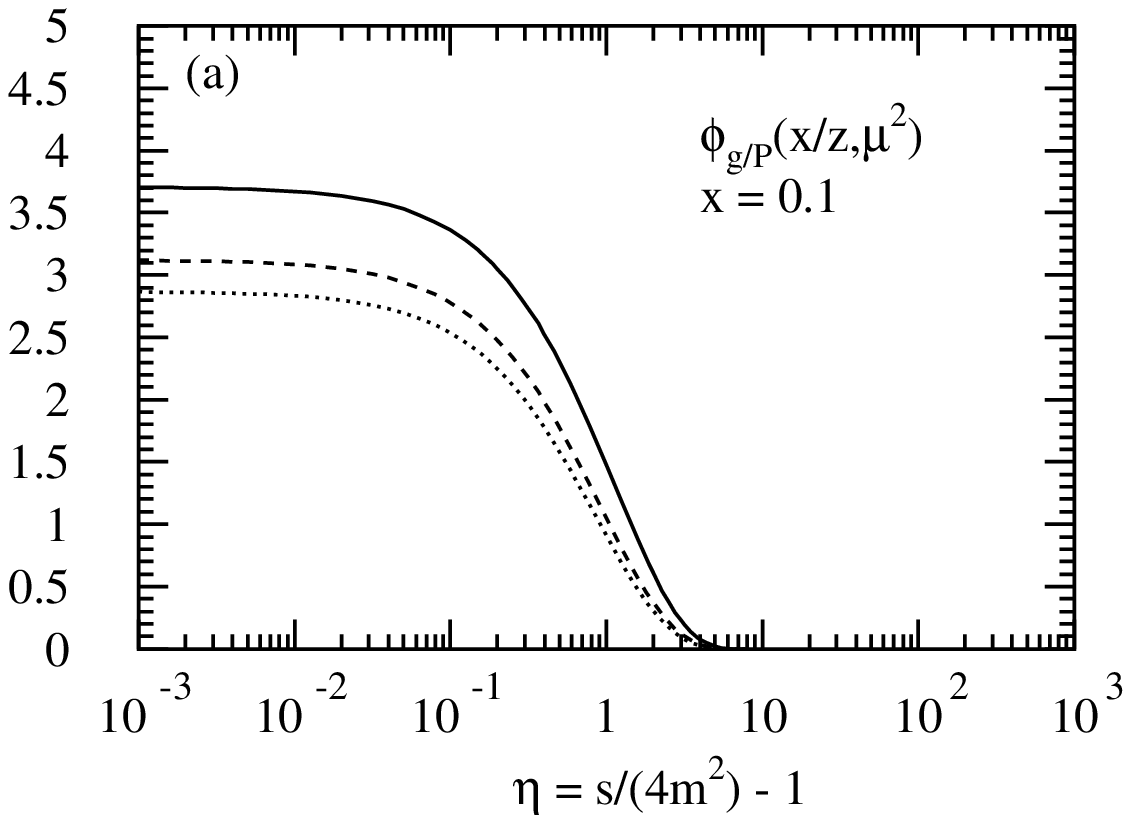,%
bbllx=50pt,bblly=130pt,bburx=285pt,bbury=470pt,angle=270,width=8.25cm}
\epsfig{file=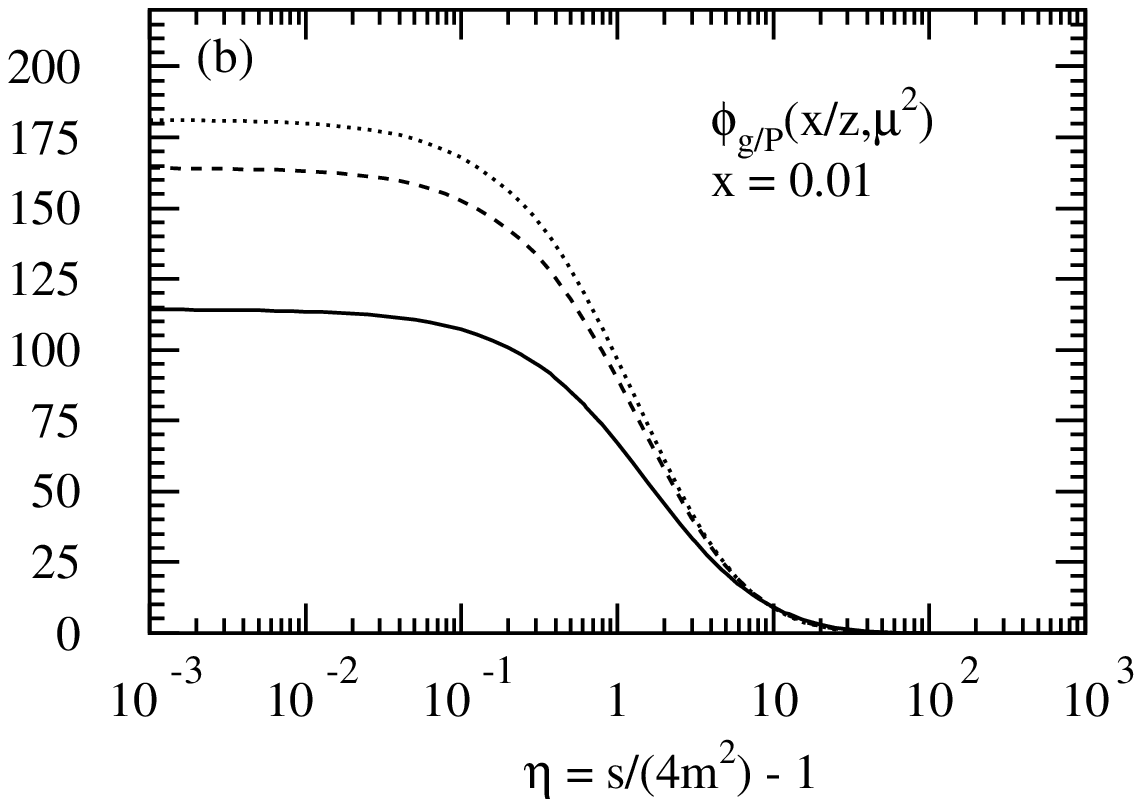,%
bbllx=50pt,bblly=110pt,bburx=285pt,bbury=450pt,angle=270,width=8.25cm}
\caption[dum]{\label{plot-pdf-1} {\small{
(a): The CTEQ4M gluon density $\f_{g/P}(x/z,\m^2)$ as a function of 
$\h$ for $m =1.6\,{\rm GeV}$, $Q^2 = 10\,{\rm GeV}$ and $x= 0.1$. 
Plotted are the scale choice $\m = m$ (solid line),  
$\m = 2 m$ (dashed line) and $\m = \sqrt{Q^2 +4 m^2}$ (dotted line).
(b): Same as Fig.~\ref{plot-pdf-1}a for $x= 0.01$. }}}
\end{center}
\end{figure}

A more revealing way to examine the support of the gluon density in
Eq.~(\ref{charmstrucintegrated}) is to plot the integral 
as a function of $z_{\rm{max}}$,
\beql
F_2^{\rm charm}(x,Q^2,m^2,z_{\rm{max}}) \!&\simeq&\!
\frac{\a_s(\m)\, e_{{\rm c}}^2 Q^2}{4 \p^2 m^2}\! 
\int\limits_{ax}^{z_{\rm{max}}}\, dz\, \phi_{g/P}(z,\mu^2)\,
\sum\limits_{k=0}^{\infty} (4 \p \a_s(\m))^k 
\sum\limits_{l=0}^{k} 
c^{(k,l)}_{2 g}(\h,\x) \ln^l\frac{\m^2}{m^2}\,. \,\,\,\,\,\,\,\,\,\,\,\,
\label{F2c_threshold_scan}
\eeql
Varying the theoretical cut-off $z_{\rm{max}}$ allows us see where the integral
(\ref{F2c_threshold_scan}) acquires most of its value.
The physical structure function corresponds 
to ${z_{\rm{max}}}=1$.

In Figs.~\ref{plot-f2c-one}a and \ref{plot-f2c-one}b we show the results for 
the numerical evaluation of Eq.~(\ref{F2c_threshold_scan}) and compare to the 
exact NLO calculation of Ref.~\cite{lrsvn93} for two values of 
$x=0.1, 0.01$ and two choices of the factorization scale 
$\m = m, \sqrt{Q^2 + 4 m^2}$.
The scale dependent terms in the NLL approximation 
to Eq.~(\ref{F2c_threshold_scan}) have been multiplied with a 
damping factor $1/\sqrt{1+\eta}$ in order to suppress spurious 
contributions from the large $\h$ region, i.e. far above
partonic threshold.{\footnote{See the discussion in 
section \ref{gluoncoefficientfunctions}. 
If we add the damping factor also to the 
scale independent terms, our NLL results at $\m =m$ are
only $5\%$ smaller.}}

Figs.~\ref{plot-f2c-one}a and \ref{plot-f2c-one}b show 
that, for the chosen kinematical range in $x$ and $Q^2$,
$F_2^{\rm charm}$ as defined by Eq.~(\ref{F2c_threshold_scan}) 
originates entirely
from that region in $\h$ in which we approximate the exact 
results very well. In words, for the values of $x$
shown, $F_2^{\rm charm}$ 
is completely determined by partonic processes close to threshold. 
We observe that the NLL approximation 
to Eq.~(\ref{F2c_threshold_scan}) is much better than the LL
one at the hadronic level as well. As we shall shortly show in  more detail, 
it is also significantly more stable under variations of the 
factorization scale.

\begin{figure}
\begin{center}
\epsfig{file=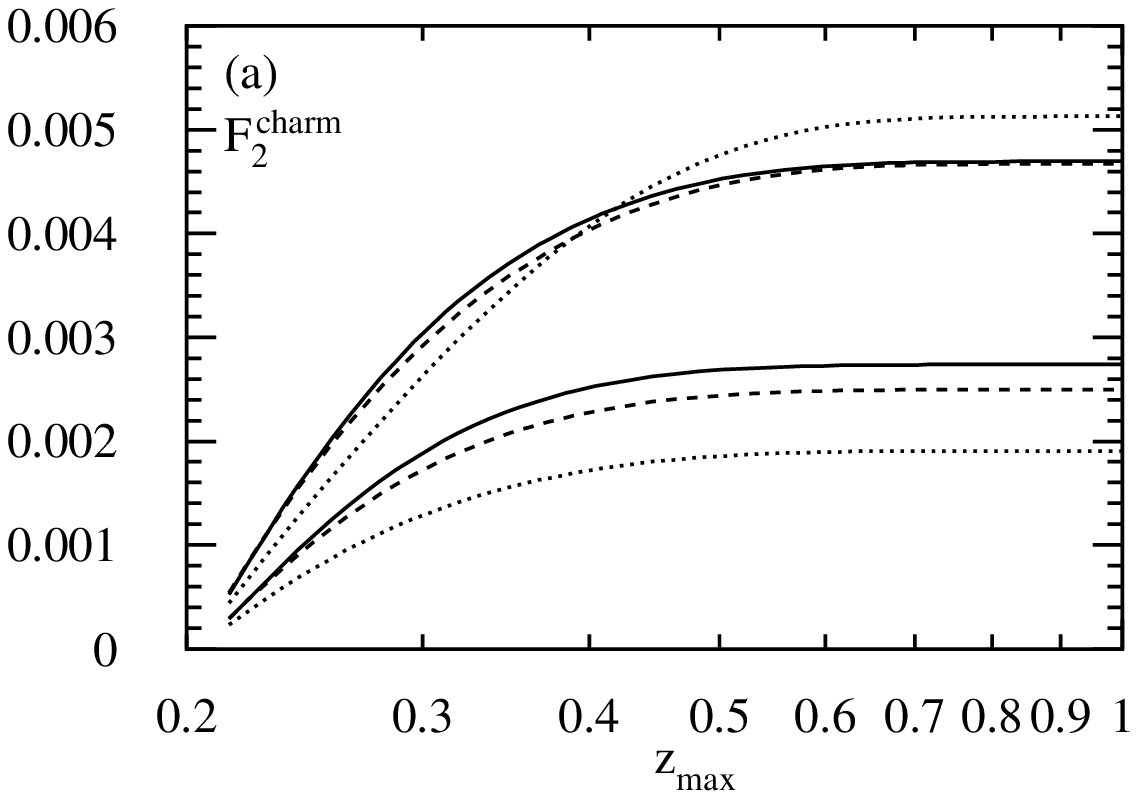,%
bbllx=50pt,bblly=130pt,bburx=285pt,bbury=470pt,angle=270,width=8.25cm}
\epsfig{file=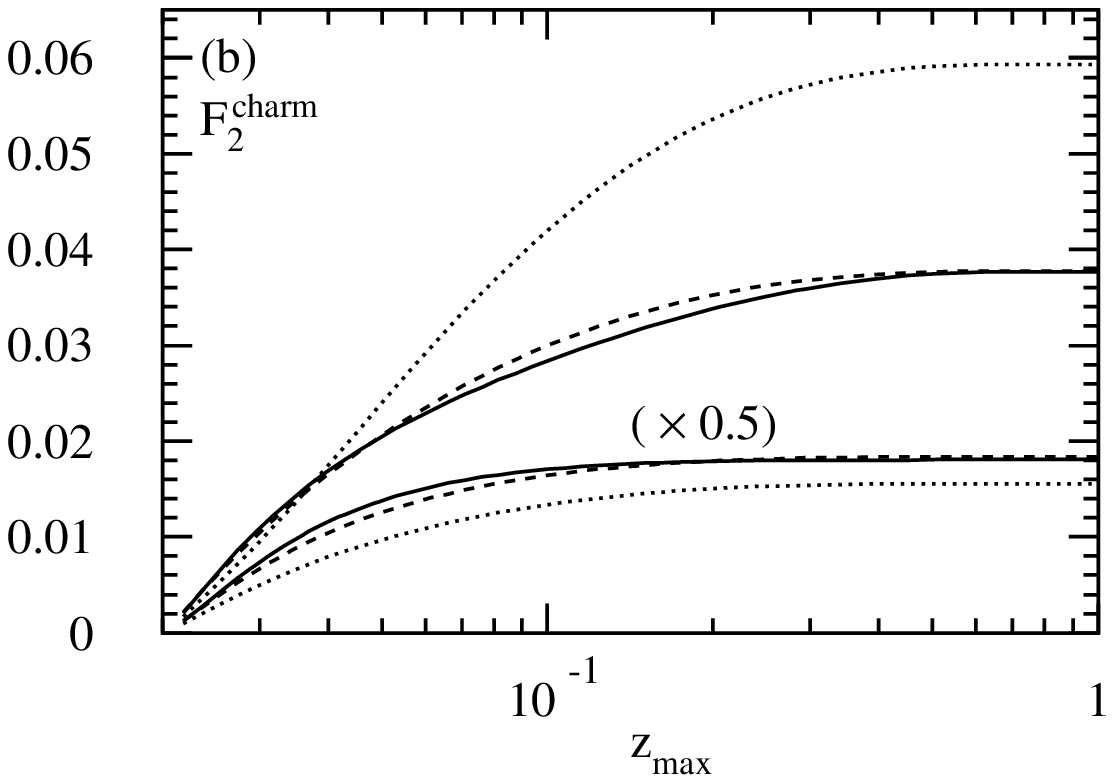,%
bbllx=50pt,bblly=110pt,bburx=285pt,bbury=450pt,angle=270,width=8.25cm}
\caption[dum]{\label{plot-f2c-one} {\small{
(a): $F_2^{\rm charm}(x,Q^2,z_{\rm{max}})$ as a function of $z_{\rm{max}}$ 
at NLO  with the CTEQ4M gluon PDF, 
$x = 0.1$, $m =1.6\,{\rm GeV}$, $Q^2 = 10\,{\rm GeV}$ 
and $\m = m$ (upper three curves), 
$\m =\sqrt{Q^2 + 4 m^2}$ (lower three curves). 
Plotted are: The exact results (solid lines), 
the LL approximations (dotted lines) and the NLL approximations 
with the damping factor $1/\sqrt{1+\eta}$ on the scale dependent terms 
only (dashed lines). 
(b): Same as Fig.~\ref{plot-f2c-one}a for $x= 0.01$. 
The lower three curves, $\m =\sqrt{Q^2 + 4 m^2}$, 
have been scaled down by a factor of 2.}}}
\end{center}
\end{figure}
\begin{figure}
\begin{center}
\epsfig{file=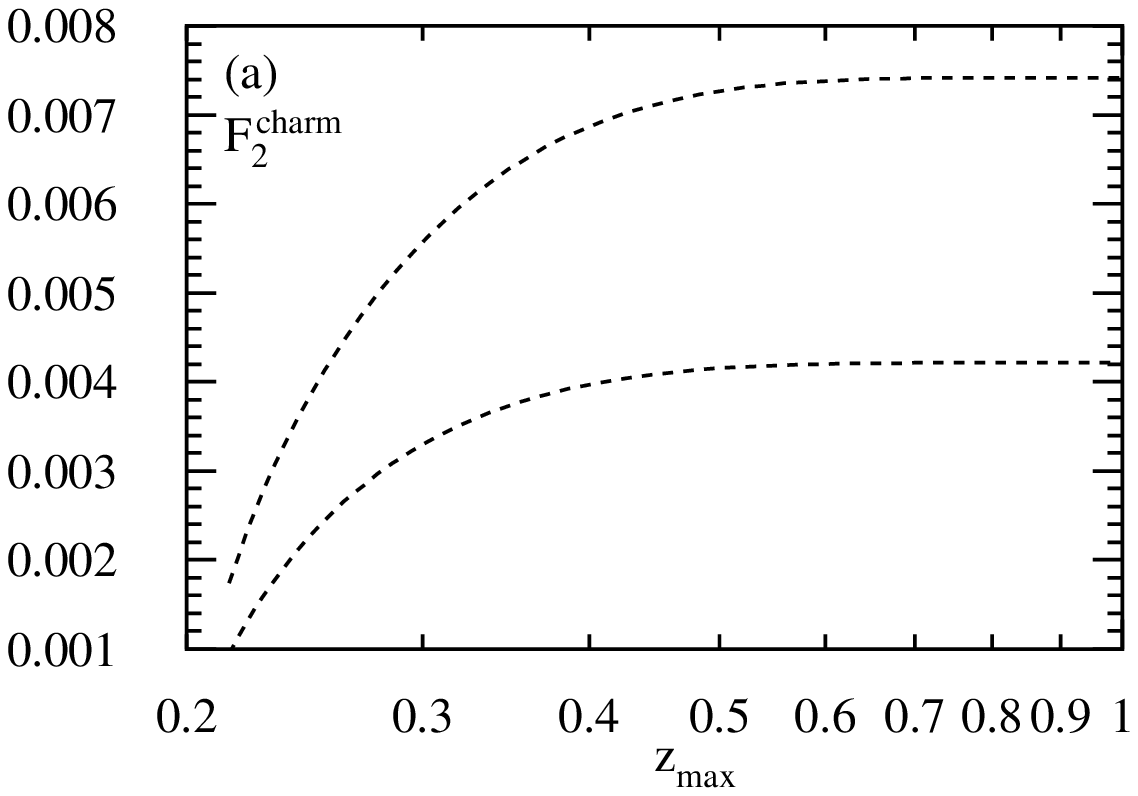,%
bbllx=50pt,bblly=130pt,bburx=285pt,bbury=470pt,angle=270,width=8.25cm}
\epsfig{file=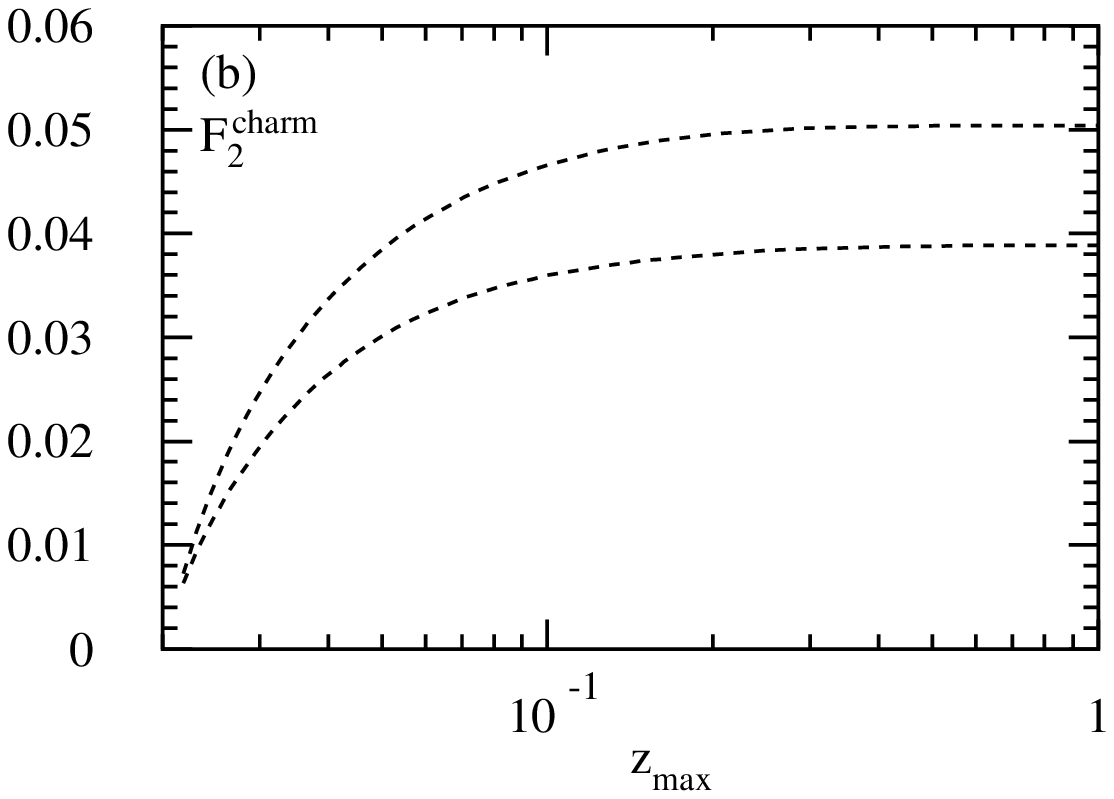,%
bbllx=50pt,bblly=110pt,bburx=285pt,bbury=450pt,angle=270,width=8.25cm}
\caption[dum]{\label{plot-f2c-two} {\small{
(a): $F_2^{\rm charm}(x,Q^2,z_{\rm{max}})$ as a function of $z_{\rm{max}}$ 
at NNLO in NLL approximation  
with the CTEQ4M gluon PDF, $x = 0.1$, $m =1.6\,{\rm GeV}$, 
$Q^2 = 10\,{\rm GeV}$ and with the damping 
factor $1/\sqrt{1+\eta}$ on all scale dependent terms and on 
$c^{(2,0)}_{2,g}$. Plotted are: Scale choices $\m = m$ (upper curve) and 
$\m =\sqrt{Q^2 + 4 m^2}$ (lower curve). 
(b): Same as Fig.~\ref{plot-f2c-two}a for $x= 0.01$.
Plotted are: Scale choices $\m = m$ (lower curve) and 
$\m =\sqrt{Q^2 + 4 m^2}$ (upper curve).}}}
\end{center}
\end{figure}

In Figs.~\ref{plot-f2c-two}a and \ref{plot-f2c-two}b we display 
the results for Eq.~(\ref{F2c_threshold_scan}) evaluated to NNLO  
in NLL approximation for the same kinematical values 
and choices for $\m$ as before. 
Again, we have multiplied the NLO scale dependent terms 
$c^{(1,1)}_{2,g}$ and also all NNLO terms
$c^{(2,l)}_{2,g}\, ,l=0,1,2$ with a damping factor $1/\sqrt{1+\eta}$.
Figs.~\ref{plot-f2c-two}a and \ref{plot-f2c-two}b reveal
that the NNLO corrections are numerically quite sizable. However, the 
variations with respect to the mass factorization scale are still 
large for both values of $x = 0.1, 0.01$.

\bigskip
\bigskip

Let us therefore turn to the issue of factorization scale dependence 
of the hadronic structure function in more detail. 
The scale dependence of $F_2^{\rm charm}$, 
cf. Eq.(\ref{charmstrucintegrated}), at NLO is exhibited in 
Figs.~\ref{plot-mu-one}a and \ref{plot-mu-one}b over a range 
in $m \le \m \le 2 \sqrt{Q^2 + 4 m^2}$. 
For $x=0.1, 0.01$
we compare the  exact result of Ref.~\cite{lrsvn93} with the LL and 
the NLL approximations from Eq.~(\ref{nloapp}). The
coefficient $c^{(1,1)}_{2,g}$ is in the approximations again
multiplied by the damping factor $1/\sqrt{1+\eta}$. 

\begin{figure}
\begin{center}
\epsfig{file=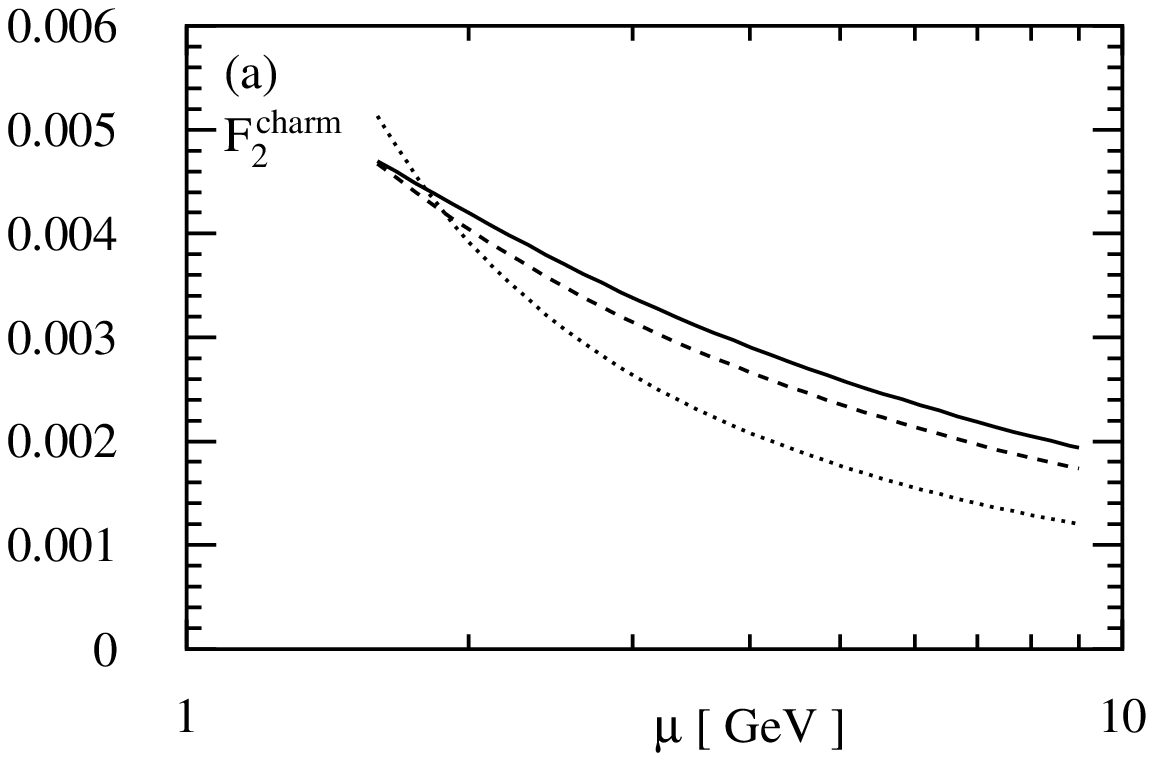,%
bbllx=50pt,bblly=130pt,bburx=285pt,bbury=470pt,angle=270,width=8.25cm}
\epsfig{file=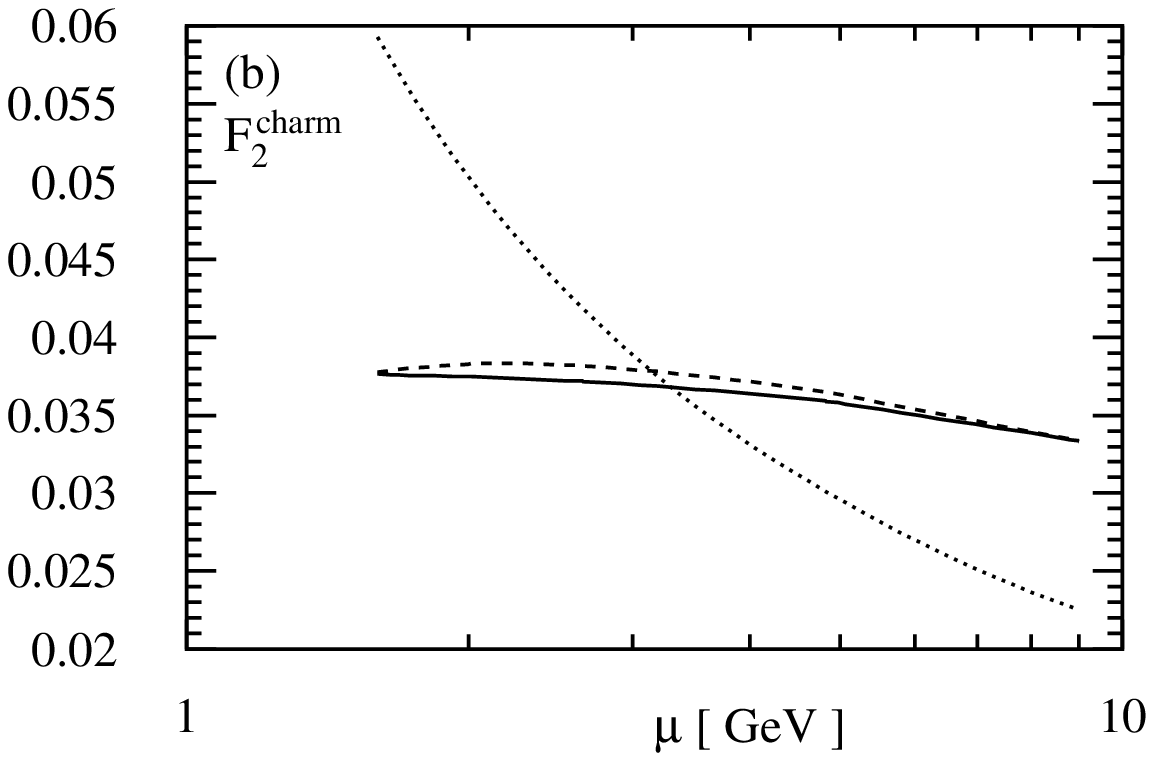,%
bbllx=50pt,bblly=110pt,bburx=285pt,bbury=450pt,angle=270,width=8.25cm}
\caption[dum]{\label{plot-mu-one}{\small{
(a): The $\m$-dependence of the charm structure function 
$F_2^{\rm charm}$ at NLO  
with the CTEQ4M gluon PDF, $x = 0.1$, $m =1.6\,{\rm GeV}$ and 
$Q^2 = 10\,{\rm GeV}$. 
Plotted are: The exact result (solid line), 
the LL approximation (dotted line) and the NLL approximation 
with the damping factor $1/\sqrt{1+\eta}$ on 
the scale dependent terms 
(dashed line). 
(b): Same as Fig.~\ref{plot-mu-one}a for $x= 0.01$.}}}
\end{center}
\end{figure}
\begin{figure}
\begin{center}
\epsfig{file=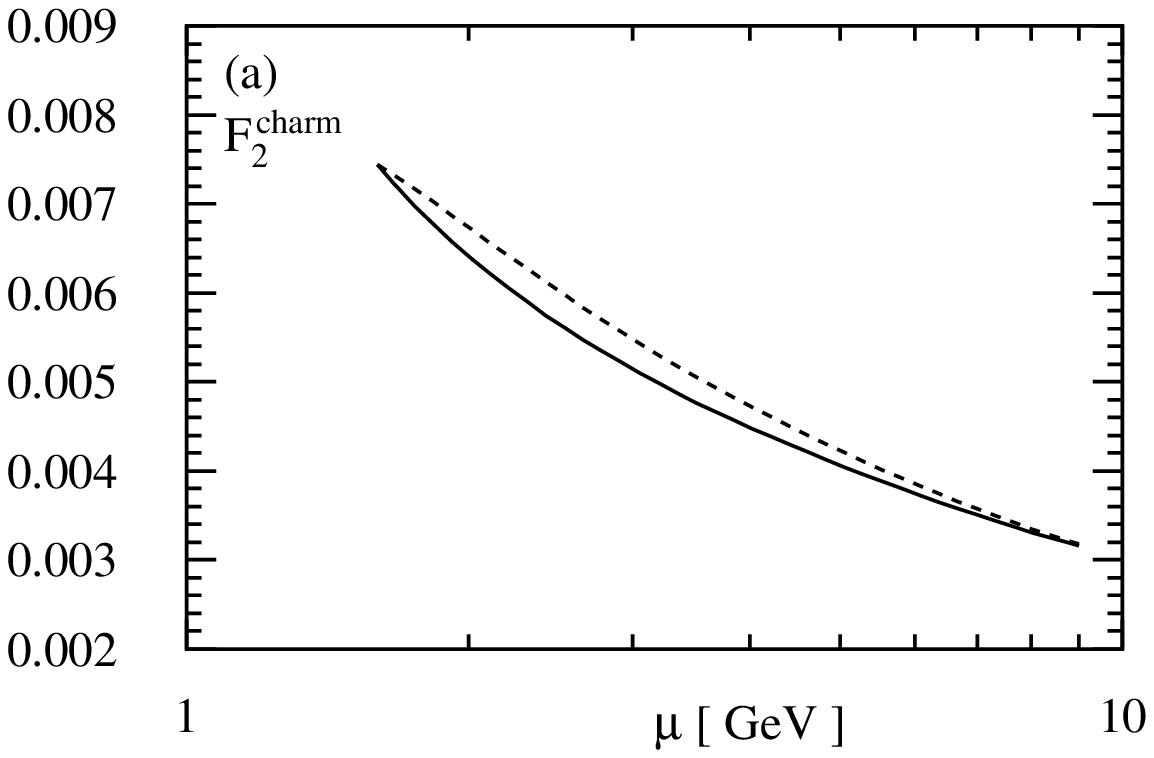,%
bbllx=50pt,bblly=130pt,bburx=285pt,bbury=470pt,angle=270,width=8.25cm}
\epsfig{file=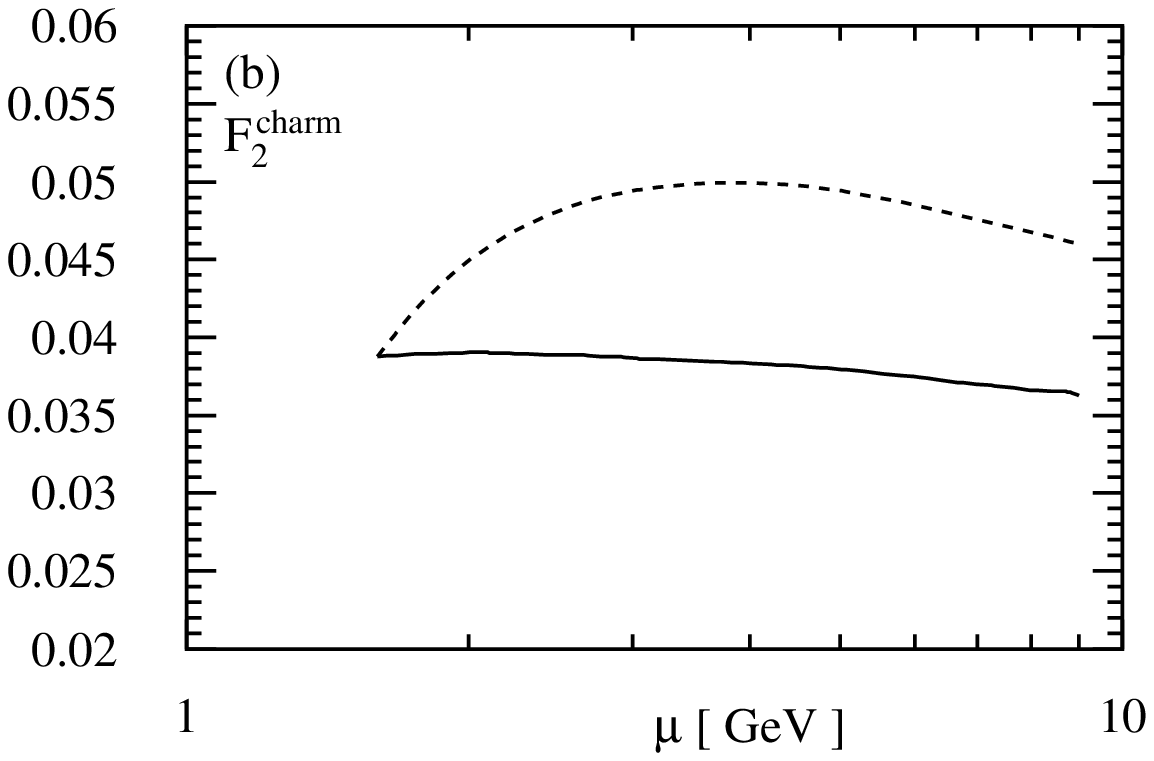,%
bbllx=50pt,bblly=110pt,bburx=285pt,bbury=450pt,angle=270,width=8.25cm}
\caption[dum]{\label{plot-mu-two}{\small{
(a): The $\m$-dependence of the charm structure function 
$F_2^{\rm charm}$ at NNLO  
with the CTEQ4M gluon PDF, $x = 0.1$, $m =1.6\,{\rm GeV}$, 
$Q^2 = 10\,{\rm GeV}$. 
Plotted are the improved NLL approximation (dashed line; exact NLO result 
plus NLL approximate NNLO result with the damping 
factor $1/\sqrt{1+\eta}$)
and the best approximation (solid line; 
exact results at NLO and for $c^{(2,1)}_{2,g}$ and $c^{(2,2)}_{2,g}$ 
plus NLL approximation $c^{(2,0)}_{2,g}$ with the damping 
factor $1/\sqrt{1+\eta}$).
(b): Same as Fig.~\ref{plot-mu-two}a for $x= 0.01$.}}}
\end{center}
\end{figure}
\begin{figure}
\begin{center}
\epsfig{file=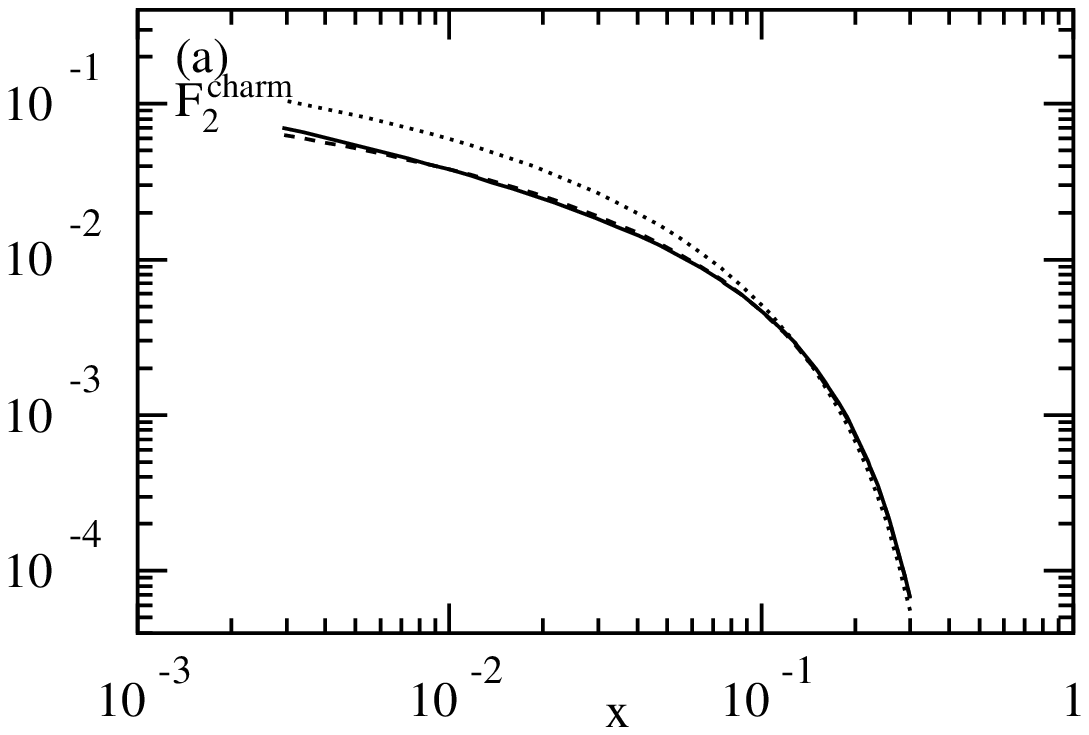,%
bbllx=50pt,bblly=130pt,bburx=285pt,bbury=470pt,angle=270,width=8.25cm}
\epsfig{file=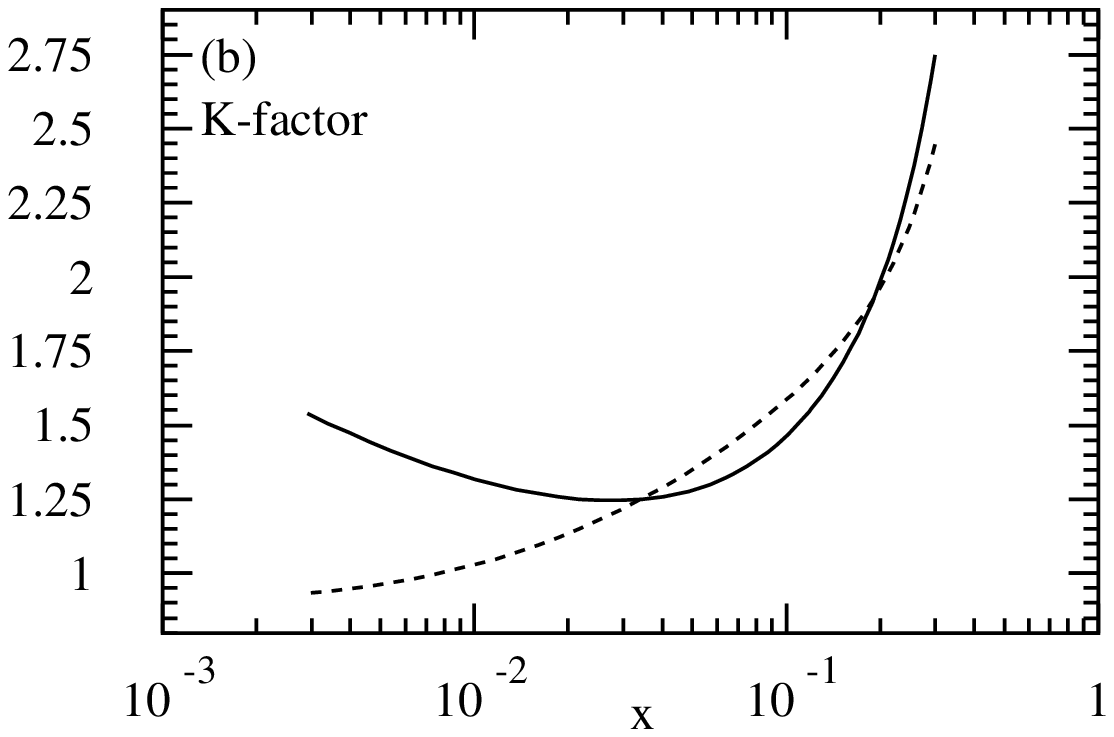,%
bbllx=50pt,bblly=110pt,bburx=285pt,bbury=450pt,angle=270,width=8.25cm}
\caption[dum]{\label{plot-xbj-one}{\small{
(a): The $x$-dependence of the charm structure function 
$F_2^{\rm charm}$ at NLO  
with the CTEQ4M gluon PDF, $\m = m = 1.6\,{\rm GeV}$ 
and $Q^2 = 10\,{\rm GeV}$. Plotted are: The exact result (solid line), 
the LL approximation (dotted line) and the NLL approximation (dashed line). 
(b):  The $x$-dependence of the ratios 
$F_{2\,\, (NLO)}^{\rm charm}/F_{2\,\, (LO)}^{\rm charm}$ (solid line) and 
$F_{2\,\, (NNLO)}^{\rm charm}/F_{2\,\, (NLO)}^{\rm charm}$ (dashed line) 
with 
$F_{2\,\, (NNLO)}^{\rm charm}$ in the improved NLL approximation (exact NLO result 
plus NLL approximate NNLO result with the damping 
factor $1/\sqrt{1+\eta}$) and parameter choices as in 
Fig.~\ref{plot-xbj-one}a.}}}
\end{center}
\end{figure}
\begin{figure}
\begin{center}
\epsfig{file=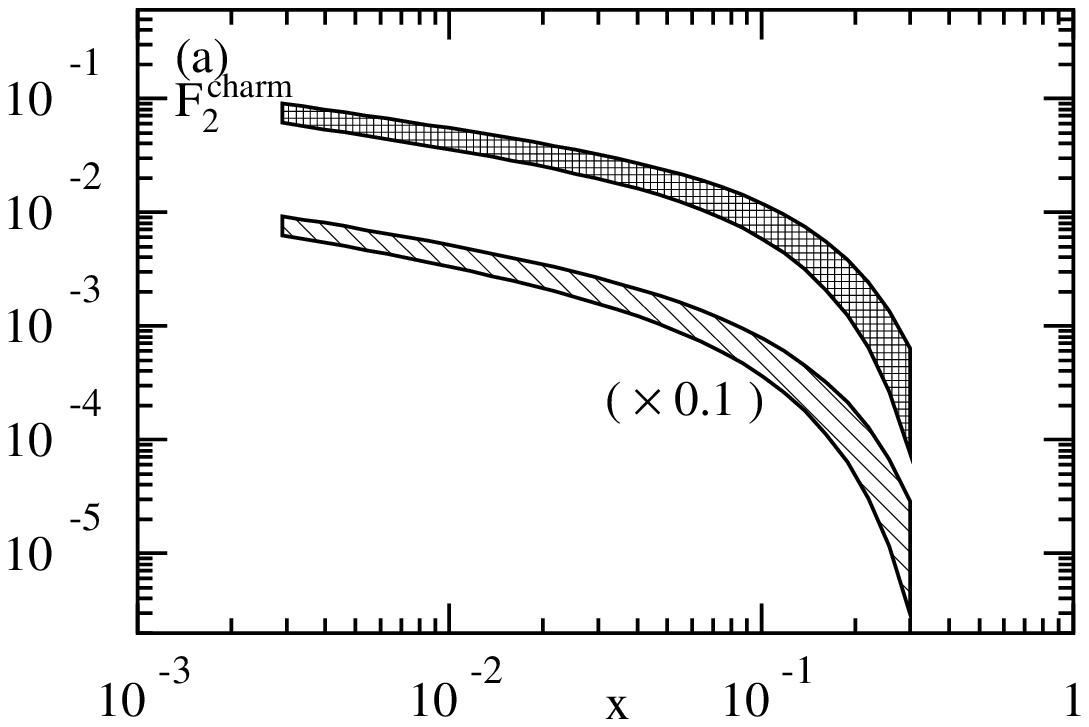,%
bbllx=50pt,bblly=130pt,bburx=285pt,bbury=470pt,angle=270,width=8.25cm}
\epsfig{file=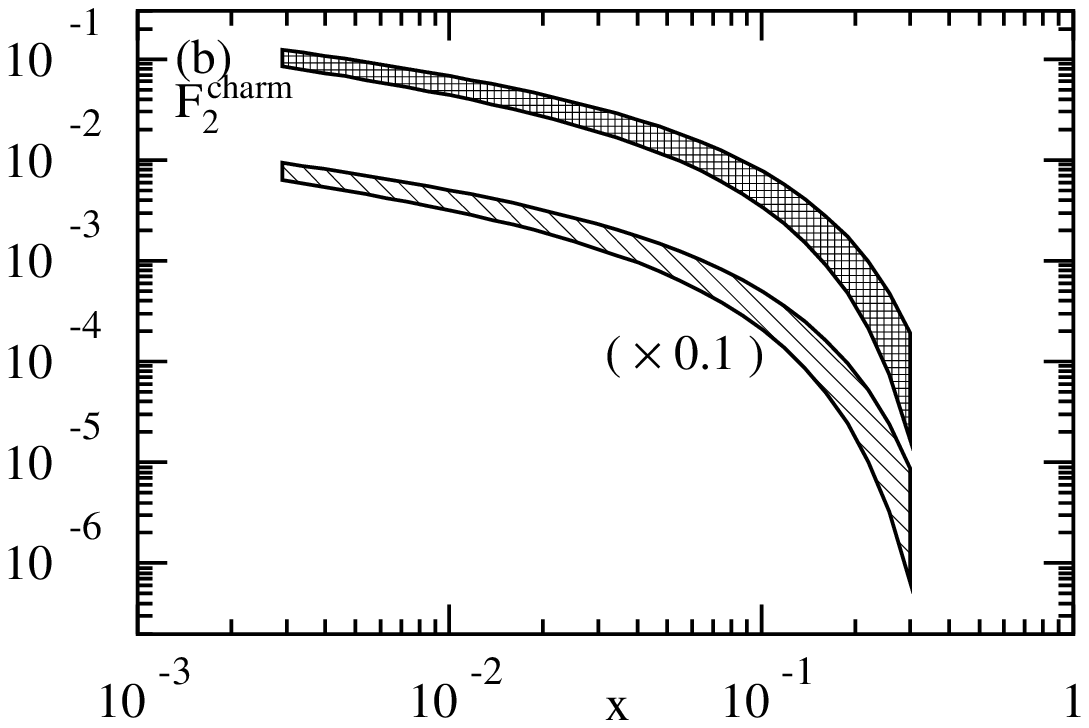,%
bbllx=50pt,bblly=110pt,bburx=285pt,bbury=450pt,angle=270,width=8.25cm}
\caption[dum]{\label{plot-xbj-two}{\small{
(a): The $x$-dependence of $F_{2}^{\rm charm}$ at NLO (lower band),
scaled down by a factor of 10 
and (upper band) in the improved NLL approximation 
at NNLO (exact NLO result plus NLL approximate NNLO result with the 
damping factor $1/\sqrt{1+\eta}$) 
 with the CTEQ4M gluon PDF, 
$\m =m$, $Q^2 = 10\,{\rm GeV}$. The bands correspond to the
charm mass variation
$1.35 \,{\rm GeV} \le m \le 1.7 \,{\rm GeV}$.
(b): Same as Fig.~\ref{plot-f2c-one}a for $\m = \sqrt{Q^2 + 4m^2}$.}}}
\end{center}
\end{figure}

We see from Figs.~\ref{plot-mu-one}a and \ref{plot-mu-one}b that 
resummation to NLL accuracy is crucial to obtain a reliable 
behaviour with respect to the $\m$-dependence of $F_2^{\rm charm}$ at NLO. 
The NLL curve traces the exact result extremely well, 
in particular at phenomenological interesting  smaller $x$ values, 
when the support of gluon PDF extends to regions of $\h = 1$ to $10$. 
In fact, Fig.~\ref{plot-mu-one}b clearly shows that in this kinematical 
domain, the LL approximation is rather poorly behaved.

Continuing, we display in Figs.~\ref{plot-mu-two}a and \ref{plot-mu-two}b 
the effect of the NNLO corrections 
on the scale dependence of $F_2^{\rm charm}$. 
We compare an {\it improved} NLL approximation to $F_2^{\rm charm}$ at NNLO,  
which consists of the exact result of Ref.~\cite{lrsvn93} 
for $F_2^{\rm charm}$ at NLO and, additionally, our NLL approximate 
NNLO results from Eq.(\ref{dsigtwoloop}), with the so-called {\it{best}} 
approximation to $F_2^{\rm charm}$ at NNLO.  
This best approximation extends beyond the NLL accuracy at two-loop order 
as it contains the full {\it{exact}} results 
for $c^{(2,1)}_{2,g}$ and  $c^{(2,2)}_{2,g}$ 
from Eqs.~(\ref{ex-c21}), (\ref{ex-c22}), such that only 
$c^{(2,0)}_{2,g}$ remains approximate to NLL accuracy.

We observe practically no improvement, with respect 
to the NLO result, in the scale dependence of $F_2^{\rm charm}$ 
at large $x$ near threshold (Fig.~\ref{plot-mu-two}a), 
where the relative variations with respect to $\m$ were largest before. 
We also see here that the two approximations to $F_2^{\rm charm}$ at NNLO, 
the NLL improved and the best, agree very well. 
They are differing only by terms beyond the NLL accuracy, 
which turn out to be rather small numerically. 
Therefore it is also unlikely that any corrections 
to $c^{(2,0)}_{2,g}$ beyond the NLL accuracy will help to  
soften the large $\m$-dependence of $F_2^{\rm charm}$ at large $x$.

On the other hand, for smaller $x$ (Fig.~\ref{plot-mu-two}b), we 
indeed see an indication of increased NNLO stability. 
The best NNLO approximation 
to $F_2^{\rm charm}$ reduces the value of the relative variation 
$F_2^{\rm charm}(\m=m)/F_2^{\rm charm}(\m=2 \sqrt{Q^2 + 4m^2} )$ from 1.13 
at NLO to 1.07 at NNLO, so that there is 
practically no dependence on $\m$ left. 
Naturally, for smaller $x$, where one probes regions away from 
threshold, terms beyond the NLL accuracy also have some numerical importance, 
as one may see from the difference between the two approximations 
to $F_2^{\rm charm}$ in Fig.~\ref{plot-mu-two}b. Finally, we note
that a fully consistent NNLO scale dependence study would require 
not-yet-available NNLO parton densities, and the presently 
also unknown three-loop anomalous dimensions.

\bigskip
\bigskip

The final comparison in this subsection involves the size
and $x$ dependence of $F_2^{\rm charm}$ at a fixed value of the 
factorization scale $\m = m$. In Fig.~\ref{plot-xbj-one}a we compare 
the exact result of Ref.~\cite{lrsvn93} 
with the LL and the NLL approximations from Eq.~(\ref{nloapp}),
over a range of 
$x$, $0.003 \le x \le 0.3$. 
We find that the deviations of the NLL approximation from the exact 
result are often very small, at most $10 \%$ for the value of $x=0.003$. 
The excellence of the NLL approximation to the $x$ dependence
also holds for other values of $\mu$.

In Fig.~\ref{plot-xbj-one}b we display for the same 
kinematics and over the same range in $x$ the effect of the 
NNLO corrections. We plot the K-factors 
$F_{2\,\, (NNLO)}^{\rm charm}/F_{2\,\, (NLO)}^{\rm charm}$ and, 
for comparison, also  $F_{2\,\, (NLO)}^{\rm charm}/F_{2\,\, (LO)}^{\rm charm}$
\footnote{For $F_{2\,\, (LO)}^{\rm charm}$
we used a two-loop $\alpha_s$ and NLO gluon density.}.
At NNLO we have taken the improved NLL approximation to $F_2^{\rm charm}$
(we recall: the exact NLO result plus the NLL approximate NNLO result 
with the damping factor $1/\sqrt{1+\eta}$). 
We see that particularly for smaller $x$, the size of the 
NNLO corrections is negligible, the K-factor being close to one, 
whereas for larger $x$, their overall size is still quite big, 
almost a factor of 2 at $x=0.1$.  

We note furthermore that the largest error on the approximate
NNLO $F_2^{\rm charm}$, as at NLO \cite{hera1}, 
is still due
to the uncertainty in the charm mass. Varying the mass of
the charm from 1.35 to 1.7 GeV, we found sizeable variations 
in the absolute value of $F_2^{\rm charm}$, in particular a larger $x$. 
In Figs.~\ref{plot-xbj-two}a and \ref{plot-xbj-two}b we plot 
for two values of the factorization scale 
$\m$ the improved NLL 
approximation to $F_2^{\rm charm}$
as a function of $x$, $0.003 \le x \le 0.3$. For comparison,  
Figs.~\ref{plot-xbj-two}a and \ref{plot-xbj-two}b also contain the exact 
NLO result with same variations of the charm mass. In general, decreasing 
the charm mass increases $F_2^{\rm charm}$.

This concludes our investigation of the inclusive structure function 
$F_2^{\rm charm}$.

\subsection{$d F_2^{\rm charm}/ d p_T$}

In this subsection we study the distribution $d F_2^{\rm charm}/ d p_T$. 
Soft gluon resummation might be especially fruitful here
because the requirement that the detected charm quark has 
a fixed transverse momentum increases the sensitivity to
threshold dynamics by effectively increasing the heavy quark 
mass from $m$ to the transverse mass $m_T = \sqrt{m^2+p_T^2}$. 

Eq. (\ref{d2ffact}) provides us with the differential 
distribution $d^2F_2/dT_1 dU_1$ from which we can derive 
the experimentally relevant distribution $d^2F_2/dy dp_T$, with 
$y$ being the rapidity in the c.m. frame of the vitual photon-proton system 
and $p_T$ the transverse momentum of the 
detected heavy quark. In terms of the Mandelstam variables of 
Eq.~(\ref{hadronicinv}), we can write the energy 
of the outgoing heavy-quark as 
\beql
E &=& -\, \frac{Q^2 + T_1 + U_1}{2\, \sqrt{S}}\, .
\label{energyheavy-quark}
\eeql
The transverse momentum $p_T$ and the longitundinal momentum $p_L$
are determined by 
\beql
S^{\prime\, 2}(p_T^2 + m^2) &=& 
S^{\prime}T_1U_1 + Q^2 T_1^2 +  Q^2 S^{\prime} T_1\, , 
\label{ptdef}
\\[1ex]
p_L^2 &=& E^2 - m^2 - p_T^2\, , 
\label{pldef}
\eeql
where $S^{\prime} = S + Q^2$. 
The rapidity $y$ 
\beql
y &=& \frac{1}{2}\, \ln\left(\frac{E + p_L}{E - p_L} \right)\, .
\eeql
may then be expressed in terms of $(T_1,U_1)$ 
by Eqs.~(\ref{energyheavy-quark}) and (\ref{pldef}).
The transformation $(T_1,U_1) \to (p_T,y)$ is then defined.
Neglecting all light initial state quarks, one obtains 
from Eq.~(\ref{d2ffact}) the result for the 
inclusive distribution $dF_2/dp_T$, 
\beql
\frac{d^2F_2(x,p_T,Q^2,m^2)}{dp_T}
&\simeq& 
\frac{2\, p_T}{S^\prime}\,\int\limits_{y^-}^{y^+} dy 
\int\limits_{z^-}^{1}\frac{dz}{z}
\,\phi_{g/P}(z,\mu^2)\,
\omega_{2,g}\Big({x\over z},s_4,t_1,u_1,Q^2,m^2,\mu^2
\Big) \;, 
\label{df2dptintegral}
\eeql
where $0 \le p_T^2 \le S/4 - m^2$ in the physical region and
\beql
y^{\pm} &=& \pm\, 
{\rm{cosh}}^{-1}\left(\frac{\sqrt{S}}{2\, \sqrt{m_T}}\right)\, .
\eeql

\begin{figure}
\begin{center}
\epsfig{file=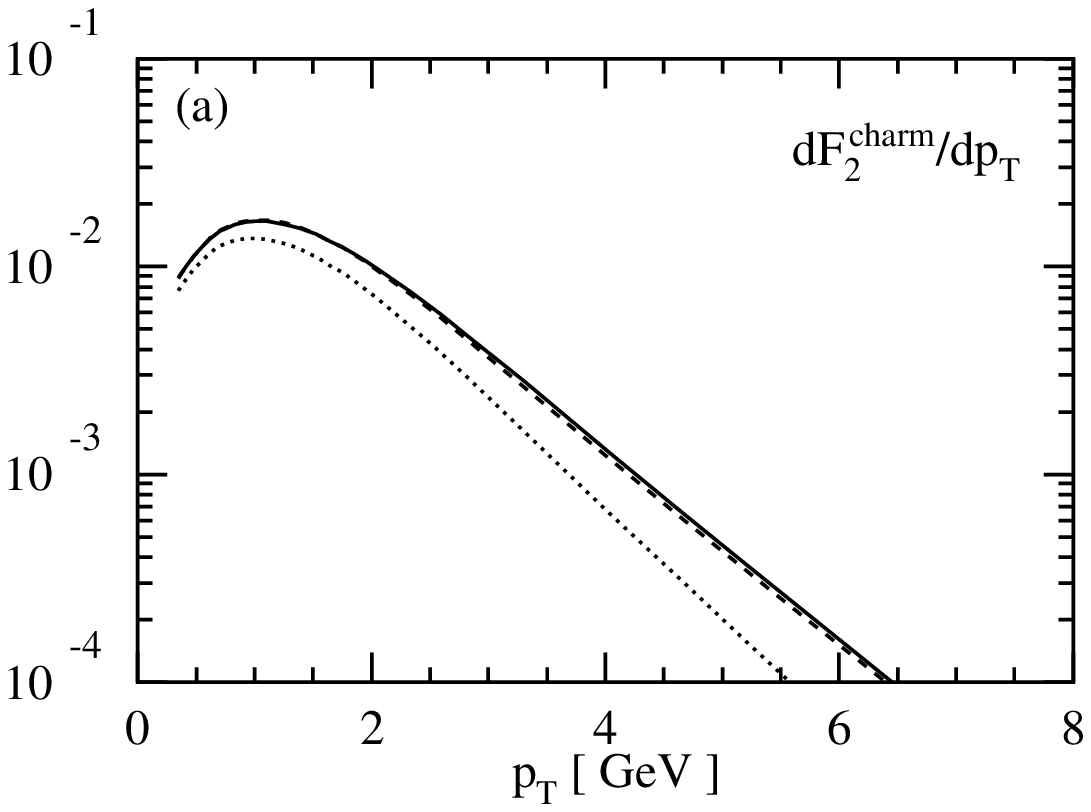,%
bbllx=50pt,bblly=130pt,bburx=285pt,bbury=470pt,angle=270,width=8.25cm}
\epsfig{file=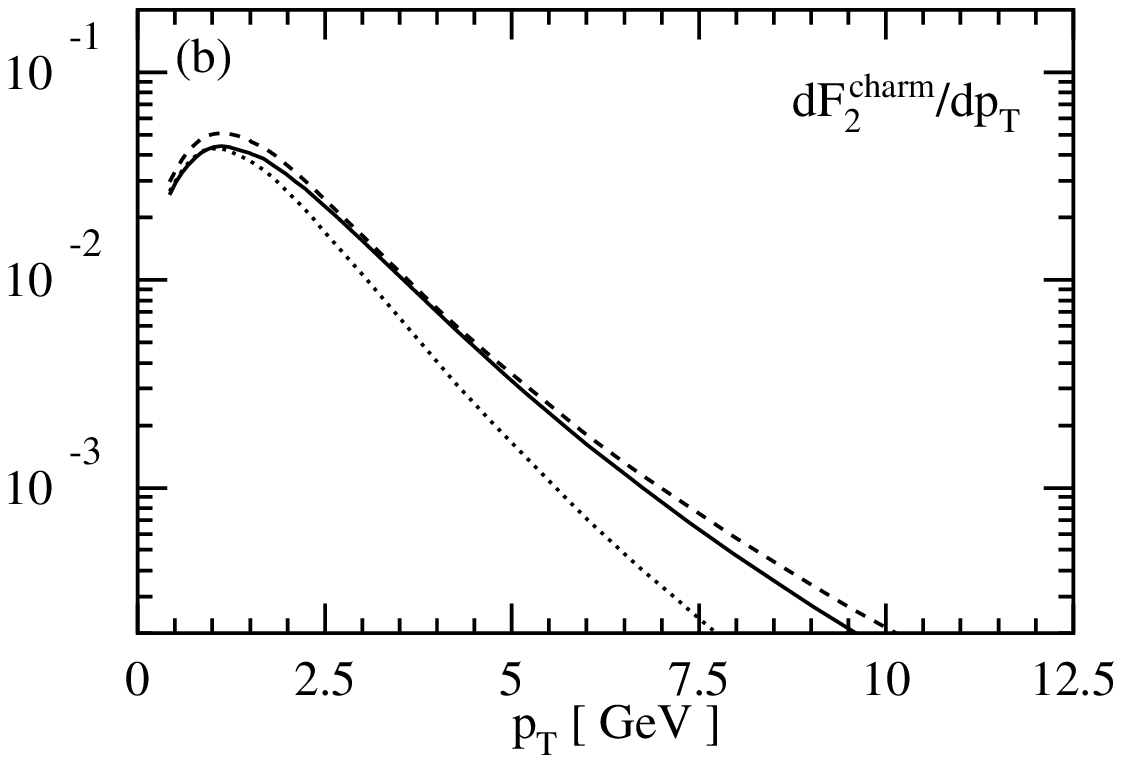,%
bbllx=50pt,bblly=110pt,bburx=285pt,bbury=450pt,angle=270,width=8.25cm}
\caption[dum]{\label{plot-pt-one}{\small{
(a): The differential distribution $d F_2^{\rm charm}/ d p_T$ 
as a function of $p_T$ at NLO  
with the CTEQ4M gluon PDF, $x = 0.01$, $m =1.6\,{\rm GeV}$, 
$Q^2 = 10\,{\rm GeV}$ and scale choice $\m = \sqrt{Q^2 + 4 (m^2 + p_T^2)}$.
Plotted are: The exact result (solid line), 
the LL approximation (dotted line) and the NLL approximation 
with the damping factor $1/\sqrt{1+\eta}$ on the scale dependent terms 
(dashed line). 
(b): Same as Fig.~\ref{plot-pt-one}a for $x= 0.001$.}}}
\end{center}
\end{figure}
\begin{figure} 
\begin{center}
\epsfig{file=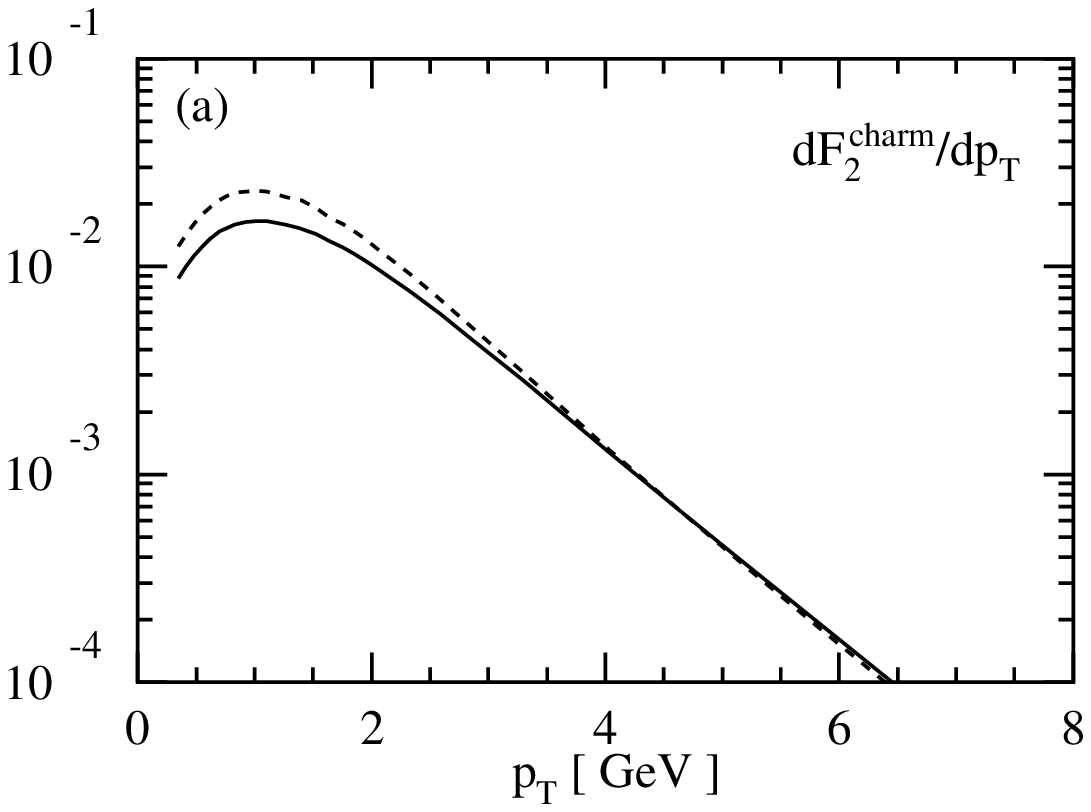,%
bbllx=50pt,bblly=130pt,bburx=285pt,bbury=470pt,angle=270,width=8.25cm}
\epsfig{file=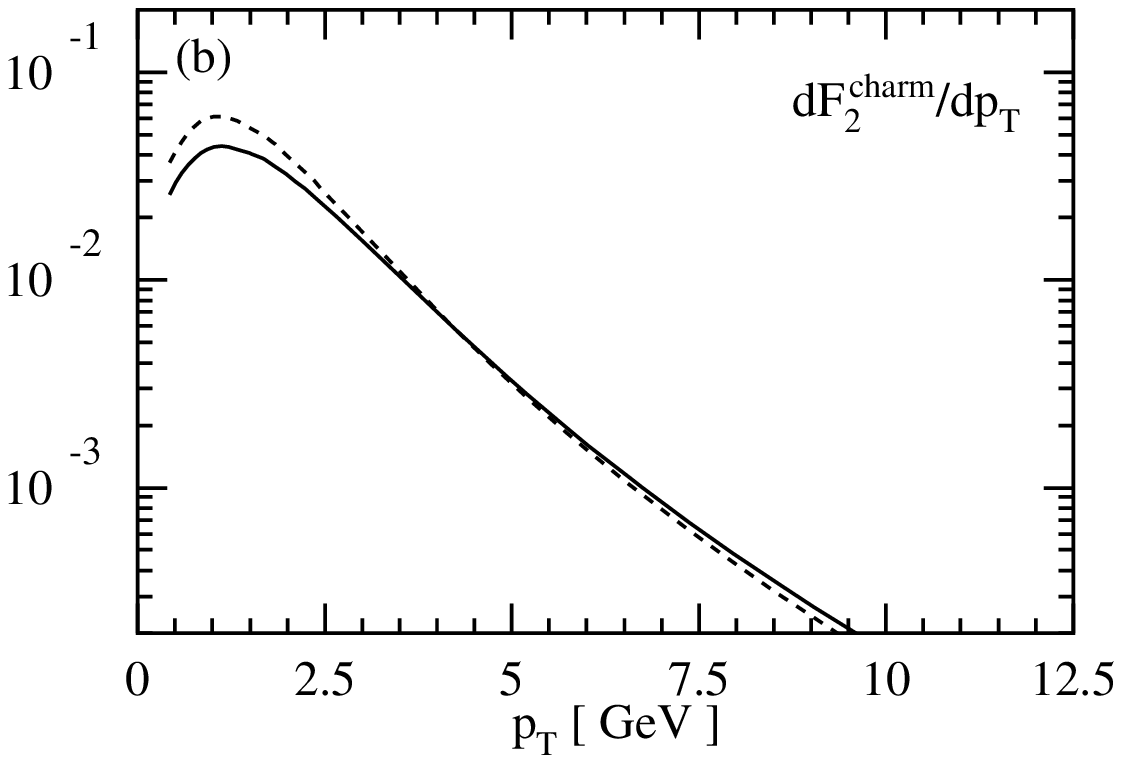,%
bbllx=50pt,bblly=110pt,bburx=285pt,bbury=450pt,angle=270,width=8.25cm}
\caption[dum]{\label{plot-pt-two}{\small{
(a): The differential distribution $d F_2^{\rm charm}/ d p_T$ 
as a function of $p_T$ with the CTEQ4M gluon PDF, $x = 0.01$, $m =1.6\,{\rm GeV}$, 
$Q^2 = 10\,{\rm GeV}$ and scale choice $\m = \sqrt{Q^2 + 4 (m^2 + p_T^2)}$.
Plotted are: The exact NLO result (solid line) and the improved NLL approximation 
at NNLO (dashed line) (exact NLO result plus NLL approximate NNLO result with the 
damping factor $1/\sqrt{1+\eta}$).
(b): Same as Fig.~\ref{plot-pt-two}a for $x= 0.001$.
}}}
\end{center}
\end{figure}

With the above expressions, and using the one- and two-loop
expansions of the hard part in section 3.1, we can construct
NLL approximations to $d F_2^{\rm charm}/ d p_T$ at NLO and
NNLO. In analogy to the inclusive $F_2^{\rm charm}$, we can then 
compare the approximate NLO results with the exact ones 
of Ref.~\cite{lrsvn93PTdistr}, and, where appropiate, make NNLO estimates.

In Figs.~\ref{plot-pt-one}a and \ref{plot-pt-one}b 
we show the NLO results vs. $p_T$ for $x=0.01$ and $0.001$, and  
$m =1.6\,{\rm GeV}$ and $Q^2 = 10\,{\rm GeV}$. 
As in Ref.~\cite{lrsvn93PTdistr}, the factorization scale has 
been chosen as $\m = \sqrt{Q^2 + 4 (m^2 + p_T^2)}$.{\footnote{
We have checked that the distribution 
$d F_2^{\rm charm}/ d p_T$, when integrated over $p_T$,
reproduces the inclusive charm structure function  $F_2^{\rm charm}$ 
in section \ref{charmstructurefunction} for the choice $\m = \sqrt{Q^2 + 4 m^2}$.}}

At NLO, we compare our LL and NLL approximate results with the 
exact results of Ref.~\cite{lrsvn93PTdistr} and we
see that the exact curves are reproduced well 
both in shape and magnitude by our NLL approximations, 
whereas the curves for LL accuracy systematically underestimate
the true result. In particular, we want to point out that for 
$d F_2^{\rm charm}/ d p_T$ the kinematical region in which our NLL 
threshold approximation applies, extends well down to $x=0.001$, with an 
error of at most $15\%$ for small $p_T$, see Fig.~\ref{plot-pt-one}b.
The deviation at larger $p_T$ is presumably related to $\ln(p_T/m)$
logarithms, associated with heavy quark fragmentation.

For the same choice of kinematics, 
Figs.~\ref{plot-pt-two}a and \ref{plot-pt-two}b display the results 
for the improved NLL approximation to $d F_2^{\rm charm}/ d p_T$ 
at NNLO. We see that the 
small $p_T$ region receives sizeable contributions, the value 
of the maximum increases by $40\%$ - $50\%$, 
whereas in the large $p_T$ region there is little change.

\section{Conclusions}

We have performed the resummation of threshold logarithms
for the electroproduction of heavy quarks, to next-to-leading
logarithmic accuracy, and in single-particle inclusive 
and pair invariant mass kinematics.
For the former, we have executed an extensive numerical investigation 
into the quality of the approximation for the inclusive charm structure
function, and for its transverse momentum distribution.
In addition we have provided both analytical and
numerical results for NNLO approximations, for those kinematic
configurations where the NLO approximation fits the exact
results well. 

We found that
the region of applicability of our analysis extends well into the 
kinematic range of the HERA experiments, a result of 
the reaction being largely driven by initial state 
gluons, which on average have not much energy.
Our studies show a clear superiority of {\it next-to-}leading logarithmic
threshold resummation over leading logarithmic resummation.
In view of the present
and much larger future HERA data sample, 
we hope that our results will find use, for example in the determination of
the gluon density, and its uncertainty, and, more generally,
that they provide motivation for further application and development 
of threshold resummation techniques.

\subsection*{Acknowledgments}

We would like to thank Gianluca Oderda, Jack Smith and George
Sterman for enlightning discussions.
This work is part of the research program of the 
Foundation for Fundamental Research of Matter (FOM) and 
the National Organization for Scientific Research (NWO).

\def\appendix{\setcounter{section}{0}
              \setcounter{equation}{0}
 \def\thesection{Appendix \Alph{section}:}
 \def\theequation{\Alph{section}.\arabic{equation}}}
\appendix

\section{Laplace transforms}
\newcommand{\Nt}{\tilde{N}}
\newcommand{\lnNt}{\ln\tilde{N}}

In this appendix we list the Laplace transform table 
needed to obtains the results of section 3. 
Define
\beq
I_n(N) = \int\limits_0^\infty~d ~    
\left(\frac{s_4}{m^2}\right)\, e^{-N\,s_4/m^2}
\left[{\ln^n(s_4/m^2)\over s_4/m^2}\right]_+\, .
\eeq
For the lowest four values of $n$ this integral is, up
to $O(1/N)$
\beql
I_0(N) &=& -\lnNt  \label{i0}\, ,\\
I_1(N) &=& \frac{1}{2}\ln^2\Nt + \frac{1}{2}\zeta_2\, ,  \\
I_2(N) &=& -\frac{1}{3}\ln^3\Nt -\zeta_2\lnNt -\frac{2}{3}\zeta_3\, ,\\
I_3(N) &=& \frac{1}{4}\ln^4\Nt + \frac{3}{2}\zeta_2\ln^2\Nt
 + 2\zeta_3\lnNt +\frac{3}{4}\zeta_2^2+\frac{3}{2}\zeta_4 \, ,
\eeql
with $\Nt = N e^{\gamma_E}$ and $\gamma_E$ denoting the Euler constant.
Note that these answers are identical to the ones for the Mellin
transform in \cite{ct}.

\setcounter{equation}{0}
\section{NLL resummed heavy quark pair-inclusive cross section}

Here we consider the singular behavior of the 
cross section for heavy
quark electroproduction, in pair-invariant mass (PIM) kinematics.
We present the resummed cross section, and its one- and
two-loop expansions.
Specifically, we consider the reaction
\beql
\g^*(q) +\, P(p) &\longrightarrow& 
{\rm{Q}{\Bar{\rm{Q}}}}(p') +\, X(p_X)\, ,
\label{photon-gluon-fusion-pim}
\eeql
We denote $p'^2=(p_1+p_2)^2=M^2$ and define $\tau = M^2/S$.
The definitions of other invariants are given in section 2.
The cross section of interest is differential with respect to $M^2$,
the scattering angle $\theta$ (in the pair center of mass frame),
and the rapidity $y$ of the pair. 
Adopting the same approximations as in section 2, it reads
\beql
\frac{d^3\s(x,M^2,\theta,y,Q^2)}{d M^2\,d \cos\theta\, d y}
&=& {1\over S'^2} \int dz \int\frac{dx'}{x'}
\,\phi_{g/P}(x',\mu^2)\; \delta\left(y-\frac{1}{2}\ln{1\over x'}\right)\;
\delta\left(z-{x 4m^2/Q^2\over x'-x}\right) \NO \\
&\times & \omega\Big(z,\theta,{Q^2\over\mu^2},
                         {M^2\over\mu^2}, \alpha_s(\mu)
\Big) \;.
\label{dsigpim}
\eeql
We have used here the observation \cite{ls} that the inclusive
partonic cross section may be used to compute the singular
behavior of the hadronic cross section at fixed (small) rapidity.

The kinematics of the reaction (\ref{photon-gluon-fusion-pim})
near threshold is determined by the vector $\zeta^\mu = (1,\vec{0})$. 
(Recall that for 1PI kinematics $\zeta^\mu=p_2^\mu/m$).
Near threshold it decomposes as 
\beql
w=(1-\tau)&=& (1-x')\left({1\over 1-x}\right) + \frac{2k_S\cdot\zeta}{\sqrt{S}}\NO \\
&\equiv & w_1\left({1\over 1-x}\right) + w_S\, .
\label{threshkinpim}
\eeql
Following the methods described in section 2, we find, in moment space
\beql
\tilde{\omega}
\left( N, \theta,{Q^2\over M^2},
                         \alpha_s(M)\right)
&=&
H\left(\theta,{Q^2\over M^2},
                          \alpha_s(M)\right)\;
\tilde{S}\left(1,\alpha_s\left({M\over N}\right)\right)
\nonumber \\
&\ & \hspace{-50mm}\times\;
\exp \Bigg \{ E_{(g)}\left(N \left(\frac{1}{1-x}\right),M^2\right) \Bigg \}
\;
\exp\Bigg \{\int\limits_{M}^{{{{M/N}}}}{d\mu'\over\mu'} 
2\, {\rm Re} \left\{\Gamma_S\left(\alpha_s(\m^{\prime})\right)\right\}\Bigg\}\, .
\label{sigNHSfinalpim}
\eeql
For simplicity, we have set the factorization scale $\m = M$, however 
it may be easily restored following the arguments leading to Eq.~(\ref{sigNHSfinal}).
The various functions in this expression are given explicitly
in Eqs.~(\ref{omegaexp})-(\ref{g1def}) and (\ref{softad}), in which
\beq
u_1 = -s'(1+\beta\cos\theta)/2,\quad\quad t_1 = -s'(1-\beta\cos\theta)/2\,,
\eeq
\bigskip
with $\beta = \sqrt{1-4m^2/s}$. The lowest order hard part in
Eq.~(\ref{dsigpim}), defined by
\beq
\omega^{(0)}(s',M,\theta)  = s'\,
\d(1-\tau)\, \s^{\rm Born}(s^{\prime}, M, \theta)\, ,
\eeq
is related to its counterpart for 1PI kinematics
in Eq.~(\ref{bornfunc1pi}) by
\beq
\label{bornpim}
\sigma^{\rm Born}(s^{\prime}, M,\theta)
= {2\pi^2\alpha \beta \over s' Q^2}
\sigma^{\rm Born}_{2,g}\left(s^{\prime},t_1 = {-s'\over 2}(1-\beta\cos\theta), 
u_1 = -{s'\over 2}(1+\beta\cos\theta)\right)\, .
\eeq
The NLL approximation to the exact NLO correction with $\m = M$ is 
\beql
\omega^{(1)}(s',\theta,M) ~\simeq~
K^{(1)}\,\s^{\rm Born}(s^{\prime}, \theta, M)\, ,
\eeql
where
\beql
K^{(1)} &=& 
 \frac{\a_s(M)}{\p}\,
\Biggl[\,
2\, C_A\, \left[{\ln(1-z)\over 1-z}\right]_+ +\,
 \left[{1 \over 1-z}\right]_+
\Biggl\{ C_A \left( -2
\ln\left( {1\over 1-x} \right) + {\rm{Re}}L_\b 
\right.
\nonumber \\[1ex]
& &\hspace*{0mm}
+ \left.\ln\left({t_1 u_1 \over m^4} \right)
- \ln\left({M^2 \over m^2} \right) \right)
- 2\, C_F \left( {\rm{Re}}L_\b + 1 \right)
\Biggr\}\Biggr] \,,
\label{nloapppim}
\eeql
where
\beq
\left[{\ln^{l}(1-z)\over 1-z}\right]_+
= \lim_{\delta \rightarrow 0} \Bigg\{
{\ln^{l}(1-z)\over 1-z} \theta(1-z -\delta)
+ \frac{1}{l+1}\ln^{l+1}(\delta)\, \delta(1-z) \Bigg\}
\label{1mzdistdef}\,.
\eeq
The NNLO corrections with $\m = M$ are to NLL accuracy 
\beql
 \omega^{(2)}(s',\theta,M)~\simeq~
K^{(2)}\,\s^{\rm Born}(s^{\prime}, \theta, M)\,,
\label{dsigonelooppim}
\eeql
where
\beql
\displaystyle
K^{(2)}&=& \frac{\a^2_s(M)}{\p^2}\,
\Biggl[\,
2\, C_A^2\, \left[{\ln^3(1-z)\over 1-z}\right]_+
+\, \left[{\ln^2(1-z)\over 1-z}\right]_+
\Biggl\{  3\, C_A^2\,  \Biggl( 
\ln\left( \frac{t_1 u_1}{m^4} \right) + {\rm{Re}}L_\b 
\NO \\[1ex]
& &\hspace*{4mm}
\displaystyle
-6 \ln\left( {1\over 1-x} \right) 
-3 \ln\left( \frac{M^2}{m^2} \right) \Biggr) 
 - 2\, C_A\, 
\left( b_2 + 3\, C_F \left( {\rm{Re}}L_\b + 1 \right) \right) \Biggr\} \Biggr]\, .
\label{dsigtwolooppim}
\eeql

\end{document}